\documentclass[fleqn,usenatbib]{mnras}

\usepackage{newtxtext,newtxmath}
\usepackage[T1]{fontenc}

\DeclareRobustCommand{\VAN}[3]{#2}
\let\VANthebibliography\thebibliography
\def\thebibliography{\DeclareRobustCommand{\VAN}[3]{##3}\VANthebibliography}

\usepackage{graphicx}	% Including figure files
\usepackage{amsmath}	% Advanced maths commands
\usepackage{orcidlink}

\newcommand{\CIV}{\ion{C}{iv}}

\newcommand{\BHM}{$M_\text{BH}$}
\newcommand{\lambdaEdd}{$\lambda_\text{Edd}$}

\defcitealias{hidalgo_connection_2022}{RH\&R22}
\defcitealias{rankine_placing_2021}{R21}

\newcommand{\change}[1]{#1}

\title[LOFAR properties of EHVOs]{LOFAR radio properties of quasars hosting Extremely High Velocity Outflows}

\author[A. L. Rankine and P. Rodr\'iguez Hidalgo]{
Amy L. Rankine$^{\orcidlink{0000-0002-2091-1966}}$$^{1}$\thanks{E-mail: amy.rankine@ed.ac.uk (ALR)}
and Paola Rodr\'iguez Hidalgo$^{\orcidlink{0000-0003-0677-785X}}$$^{2}$
\\
$^{1}$Institute for Astronomy, University of Edinburgh, Royal Observatory, Blackford Hill, Edinburgh EH9 3HJ, UK\\
$^{2}$Physical Sciences Division, School of STEM, University of Washington Bothell, Bothell WA, 98011, USA
}

\date{Accepted XXX. Received YYY; in original form ZZZ}

\pubyear{\the\year{}}

\begin{document}
\label{firstpage}
\pagerange{\pageref{firstpage}--\pageref{lastpage}}
\maketitle

\begin{abstract}
Quasar winds are a potential contributor to the radio emission observed in the so-called `radio-quiet’ quasars. Extremely High Velocity Outflows (EHVOs), with bulk velocities $>0.1c$, represent some of the most energetic of these winds. In this work, we investigate the radio properties of quasars hosting EHVOs in order to assess whether such high-velocity winds contribute to the observed radio emission. We cross-match the catalogue of EHVOs from the Sloan Digital Sky Survey with the high-sensitivity Low-Frequency Array (LOFAR) Two-metre Sky Survey (LoTSS) Data Release 3. We find that EHVO quasars exhibit a significantly higher radio-detection fraction of 38 per cent compared to an underlying parent quasar population at 24 per cent. By constructing control samples using nearest-neighbour matching, we show that matching the {\CIV} emission line blueshift alone is sufficient to reproduce the elevated radio-detection fraction in a global sense. However, differences still remain in the radio-detection fraction as a function of {\CIV} emission blueshift between the EHVO distribution and any matched sample, suggesting that some additional mechanism is at play. Furthermore, composite spectra reveal no significant differences in the ultraviolet absorption profiles between radio-detected and radio-undetected EHVOs, although we identify a \change{tentative} positive correlation between maximum outflow velocity and radio luminosity. Our results suggest that EHVOs likely represent a subset of the quasars which host fast winds that are observed along particular lines-of-sight, and that radio emission associated with such winds remains a plausible mechanism.

\end{abstract}


%%%%%%%%%%%%%%%%%%%%%%%%%%%%%%%%%%%%%%%%%%%%%%%%%%

%%%%%%%%%%%%%%%%% BODY OF PAPER %%%%%%%%%%%%%%%%%%

\section{Introduction}
\label{sec:intro}

In studies of the radio properties of quasars, sources are commonly divided into two classes based on the relative strength of their radio and optical emission. Objects whose radio flux densities exceed their optical emission by a factor of $\gtrsim 10$ are typically classified as `radio-loud' and have radio emission that is dominated by powerful jets \citep[e.g.,][]{Strittmatter80_radio_observations, Kellermann89_vla_observations}. The remainder -- although more prevalent -- are designated `radio-quiet'.
It remains unclear what is the driver of the radio emission in radio-quiet quasars: compact/weak jets \citep[e.g.,][]{Bicknell02_connections_between}, star formation \citep[e.g.,][]{Condon92_radio_emission}, the corona \citep[e.g.,][]{Laor08_origin_radio}, and quasar winds \citep[e.g.,][]{Zakamska14_quasar_feedback, Nims15_observational_signatures} are all viable \citep{kimball_2018_3942728,Panessa19_origin_radio}.
With the advent of sensitive radio telescopes such as the Low-Frequency Array \citep[LOFAR;][]{van_haarlem_lofar_2013}, this radio-quiet emission is more easily detected, allowing investigations into its origin out to moderate redshifts.

Some previous studies have investigated the contribution of winds to the radio emission by investigating quasars with broad absorption lines (BALs) in their spectra, a definite signature of quasar winds \citep{weymann_comparisons_1991}. According to \citet{morabito_origin_2019}, BAL quasars are more likely to be radio-detected at 144\,MHz than non-BALs but they do not observe a correlation between the balnicity index (the strength of BALs, see Section~\ref{sec:data}) and radio-loudness or radio luminosity, suggesting that the BALs and radio emission are produced by physically distinct but related phenomena. \change{At GHz frequencies, \citet{Chen26_radio_emission} observed no correlation between the radio luminosity and kinetic power of the wind in a sample of 13 BAL quasars and found their radio-loudness comparable to non-BAL quasars.} \citet{petley_how_2024} have shown that dust-reddening/colour and accretion rate may be important in the BAL and non-BAL populations, with redder \citep[see also][]{fawcett_fundamental_2022} or more highly accreting sources more likely to be radio-detected. More recently, Grieve et al. (in prep) suggest that the underlying distributions of {\CIV} blueshift (the asymmetry of the {\CIV}$\lambda1549$ emission line often associated with disc winds; e.g., \citealt{Wilkes84_studies_broad, Baldwin96_very_high, Matthews23_disc_wind}), Eddington ratio ({\lambdaEdd}) and black hole mass ({\BHM}) in the BALs compared to non-BALs are contributing factors to the increased radio-detection fraction observed in the BALs. Previous works have shown that the radio detection increases with increasing {\CIV} blueshift [\citealt{rankine_placing_2021} (\citetalias{rankine_placing_2021} hereafter); \citealt{Richards21_probing_wind}], and \citet{jackson_exploring_2026} have revealed that the radio-quiet and radio-loud quasars inhabit the three-dimensional ({\CIV} blueshift, {\lambdaEdd}, {\BHM}) space differently. These studies all point to winds as the simplest explanation for the origin of a significant fraction of the radio-quiet emission; however, this question is far from resolved.

The kinetic energy in the wind may be responsible for shocking the ISM and in turn producing radio emission via synchrotron emission \citep[e.g.,][]{Zakamska14_quasar_feedback, Nims15_observational_signatures} that contribute to the radio emission in the so-called radio-quiet quasars. Classical BAL definitions encompass winds with speeds that have been arbitrarily set to a maximum of $0.1c$. 
Extremely high velocity outflows (EHVOs) are defined as winds observed in UV/optical at speeds beyond this limit, larger than 0.1$c$ \citep{PRH_2011}. While ultra fast outflows have been found at even larger speeds \citep[e.g.,][]{tombesi_evidence_2010, matzeu_supermassive_2023}, the strength and availability of observing EHVOs at rest-frame wavelengths of $\sim$1300--1400\,{\AA} make them ideal for the study of extreme winds at large redshifts; in fact, EHVOs have been found in several quasars at redshift larger than 7 \citep{Wang18,Wang21}. Given the \change{extreme} kinetic luminosities of EHVOs, even if a small fraction couples with the gas to produce shocks [either in the ISM; e.g., \citealt{Zakamska14_quasar_feedback}; or in the broad line region (BLR); e.g., \citealt{sotomayor_nonthermal_2022}], it may be possible to detect any associated radio emission. 
\citet{PRH_2025} presented a detailed study of one EHVO quasar \change{that demonstrated the potential importance of EHVOs for quasar feedback due to its mass outflow rate and kinetic luminosity.} 

Until recently, only a handful of cases of EHVOs were known. The largest samples of EHVOs to date are the 137 cases found in the DR9 and DR16 samples (\citealt{hidalgo_survey_2020}, Candelaria-Stoner et al. in prep). Given their recent discovery, little was known about the properties \change{of their host quasars} and how their extreme energies can affect their host galaxies.
BALs and non-BALs have overlapping {\lambdaEdd}, {\BHM}, and {\CIV} \textit{emission} blueshift distributions, and {\lambdaEdd} is correlated with this {\CIV} blueshift \citep{Rankine20_bal_non-bal, Temple23_testing_agn}.
For EHVOs, we discovered in \citet[][\citetalias{hidalgo_connection_2022} hereafter]{hidalgo_connection_2022} that they show larger {\CIV} \textit{emission} blueshifts overall than BALs and non-BALs. 
EHVOs also show a larger {\lambdaEdd} for a given {\BHM} (Flores et al. in prep). 
Given that EHVOs are potentially more energetic and occupy a distinctive region of the {\CIV} blueshift, {\lambdaEdd}, {\BHM} spaces, they provide an excellent sample \change{for distinguishing wind-related radio emission from other emission mechanisms.}

\citet{hidalgo_survey_2020} reported that one out of the 40 EHVOs in the DR9 sample had a non-zero FIRST \citep[Faint Images of the Radio Sky at Twenty-Centimeters;][]{Becker95_first_survey} flux (a $\sim$10-$\sigma$ detection) reported in the SDSS DR9 catalogue \citep{Paris12_sloan_digital}. The higher sensitivity of LOFAR combined with the larger sample of EHVOs available now allows a deeper investigation of the radio properties of the quasars hosting an EHVO.

In this paper, we have cross-matched the LoTSS DR3 catalogue with the EHVO and parent sample (see Section~\ref{sec:data}) to determine their radio-detection fractions and explore the relationship between EHVO properties and the radio emission (Section~\ref{sec:results}). We discuss our results in the context of other possible radio-producing mechanisms in Section~\ref{sec:disc} before concluding in Section~\ref{sec:concl}.

Vacuum wavelengths are employed throughout the paper and we adopt a $\Lambda$CDM cosmology with $h_0 = 0.71$, $\Omega_M = 0.27$, and $\Omega_\Lambda= 0.73$ when calculating quantities such as quasar luminosities.

\section{Data}
\label{sec:data}

The parent and EHVO samples were compiled from the SDSS DR9 and DR16 catalogues by \citet{hidalgo_survey_2020} and Candelaria-Stoner et al. (in prep), respectively. The parent samples in which EHVOs were searched for were defined by limiting to spectra at $z>1.9$ and with median $S/N>10$ in the windows 1250--1400\,{\AA} and 1650--1750\,{\AA} (see the above catalogue papers for more details on the selection). The combined DR9+DR16 parent sample numbers 24\,681.
Each spectrum was normalized with respect to a power-law, producing $f(V)$, and EHVOs were then searched for using a modified balnicity index (\citealt{weymann_comparisons_1991}) designed to search for absorption between 30\,000 and 60\,000\,km\,s$^{-1}$:

\begin{equation}
    \text{BI}_\text{EHVO} = \int_{30\,000}^{60\,000} \left(1 - \frac{f(V)}{0.9}\right) C\ \text{d}V,
\end{equation}
and $C=1$ where $f(V)<0.9$ for at least 1\,000\,km\,s$^{-1}$ in the DR9 selection and 2\,000\,km\,s$^{-1}$ in DR16. Any spectra with BI$_\text{EHVO}>0$ were visually inspected and the total number of EHVOs discovered was 137.

EHVO measurements, specifically BI$_\text{EHVO}$ and maximum absorption velocity, are available from \citet{hidalgo_survey_2020} and Candelaria-Stoner (in prep). We make use of the {\CIV} emission line measurements, luminosities, and black hole masses measured from the SDSS spectrum reconstructions that are based on an Independent Component Analysis decomposition and detailed in \citet{Rankine20_bal_non-bal}. The reconstructed sample covers $1.5<z<3.5$ and the largest sample to date was compiled for the analysis of \citet{Temple23_testing_agn}.
Of the parent sample, 22\,425 have been reconstructed, 111 of which host EHVOs.  

The sample is limited further to those falling within the footprint of the third data release of the LOFAR Two-Metre Sky Survey \citep[LoTSS DR3;][]{shimwell_lofar_2026}, defined by a multi-order coverage map generated from the DR3 catalogue using \textsc{mocpy} \citep{baumann_2026_mocpy} with a maximum order of 7. Our final sample for analysis contains 21\,293 quasars, 106 of which are host to EHVOs.

The {\CIV} emission blueshift is measured non-parametrically as the wavelength which bisects the emission line flux, relative to the systemic quasar redshift. The 1350 and 3000\,{\AA} monochromatic luminosities, $L_{1350}$ and $L_{3000}$, are measured directly from the reconstructions and bolometric luminosities defined as $L_\text{bol}=3.81 L_{1350}$ \citep[from][]{richards_spectral_2006}. Due to the spectral coverage of the SDSS and BOSS spectrographs and the redshift range of our sample, only the {\CIV} emission line is present in all of our objects. Thus, we use the {\CIV} full width at half maximum (FWHM) corrected to account for the non-virial blueshifted component alongside the radius--luminosity relation of \citet{vestergaard_determining_2006} to estimate black hole masses \citep[as per][]{coatman_correcting_2017}. Where {\CIV} blueshift~$<500$\,km\,s$^{-1}$, we use the uncorrected FWHM. \change{This FWHM correction reduces the scatter between the {\CIV} masses and those derived from the Balmer lines from 0.40\,dex to 0.10\,dex in low-blueshift objects and to 0.24\,dex in high-blueshift objects \citep[see fig. 11 and section 5.3 of][]{coatman_correcting_2017}.}

The EHVO and parent sample are cross-matched with the radio positions available in the LoTSS DR3 catalogue using a radius of 3\,arcsec. Based on analysis performed by Grieve et al. (in prep) using LoTSS DR2 and the careful optical cross-match of \citet{Hardcastle23_lofar_two-metre}, the majority of our quasars are unresolved in the radio and have single-component source structures such that a simple cross-match with the radio positions is appropriate. We classify a quasar as radio-detected if its peak flux is greater than 5 times the peak flux uncertainty (uncertainty in peak Stokes I flux density per beam of the source). 5108 quasars in the parent sample and 40 EHVOs are detected in the radio. Radio luminosities at 144\,MHz are measured from the total integrated flux density, $S_{144}$, provided in the LoTSS DR3 catalogue and assuming a typical \change{optically-thin} synchrotron spectral index $\alpha=-0.7$: $L_{144}=4\pi d_L^2S_{144}(1+z)^{-\alpha-1}$\,W\,Hz$^{-1}$.

\begin{figure}
    \centering
    \includegraphics[width=\linewidth]{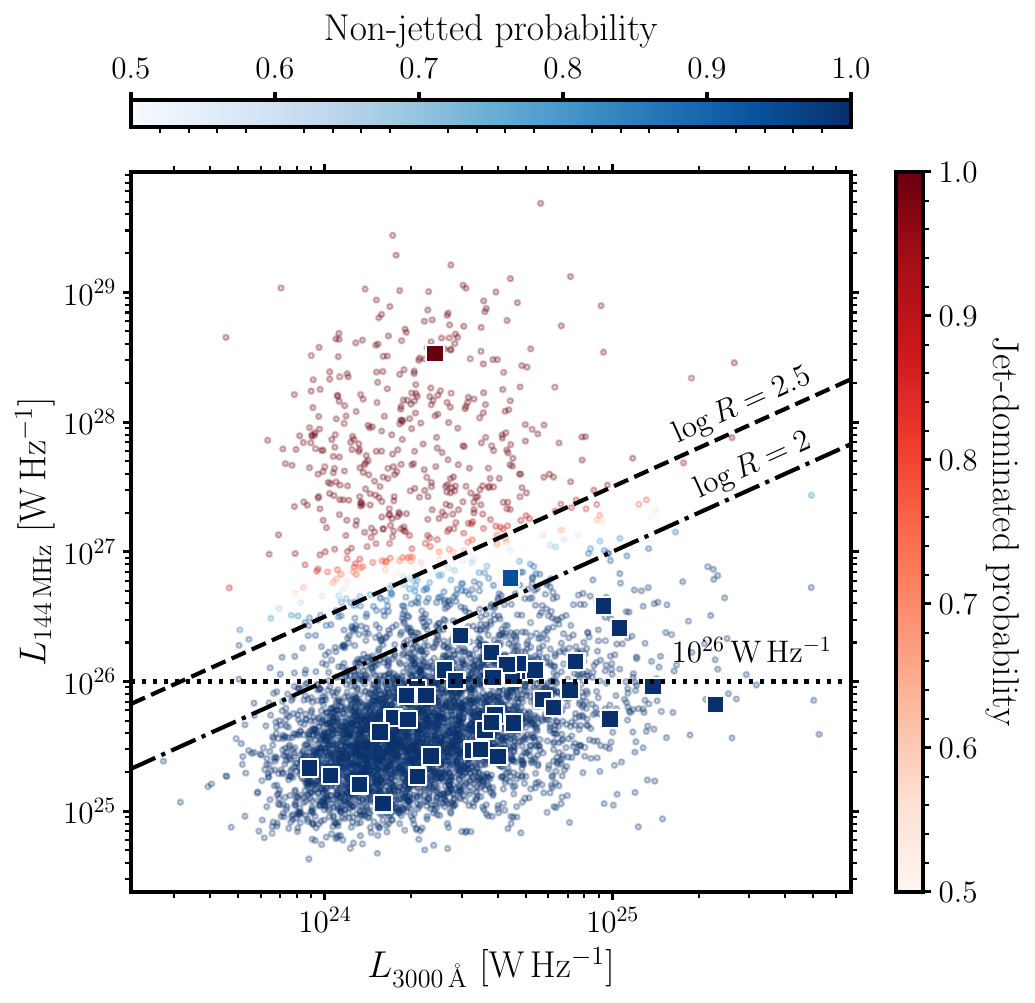}
    \caption{LOFAR radio luminosity at 144\,MHz against 3000\,{\AA} continuum luminosity. The parent sample is plotted as \change{circles and the EHVOs as squares}. Traditional cuts for separating jetted from non-jetted sources are marked: $L_{144}=10^{26}$\,W\,Hz$^{-1}$, $\log R=2$, and $\log R=2.5$ as used by \citet{jackson_exploring_2026}. We employ the same Gaussian Mixture Model technique as \citet{jackson_exploring_2026} to attempt to separate jet-dominated radio emission and colour-code the sources by the probability that they fall within the most-likely population.}
    \label{fig:lums}
\end{figure}

The radio-detected samples will include quasars with radio jets. The purpose of this analysis is to assess if the winds traced by the EHVOs are a plausible contributor to the radio emission, and as such we additionally remove sources with $\log R>2$ (i.e., radio-loud sources) to reduce the contamination of the radio emission from strong jets.
In Fig.~\ref{fig:lums} we plot the 144\,MHz radio luminosity against the 3000\,{\AA} luminosity of the parent sample and EHVOs. We translate the radio-loud threshold of $\log R=1$ at 5\,GHz to $\log R=2$ at 144\,MHz, where $R = L_{144}/L_{3000}$ (objects with $\log R < 2$ are radio-quiet) and plot this alongside $L_{144} = 10^{26}$\,W\,Hz$^{-1}$, both of which have been used in the past to demarcate radio emission dominated by strong jets. 
\citet{yue_novel_2024, yue_novel_2025} and \citet{jackson_exploring_2026} use alternative means to isolate the likely-jetted sources.
\citet{yue_novel_2024} models a two-component flux density distribution to produce the probability that a source's radio emission is dominated by jets.\footnote{We do not employ the method of \citet{yue_novel_2024} here since this would require extrapolating the model to the high optical luminosities in our sample.} \citet{jackson_exploring_2026} meanwhile makes use of a two-component Gaussian Mixture Model in $L_{3000}$--$L_{144}$ space to separate the jets and, on separate occasions, a $\log R = 2.5$ cut. We employ the same Gaussian Mixture Model and, in Fig.~\ref{fig:lums}, colour-code the samples by the probability of being correctly assigned to either the population whose radio emission is dominated by jets, or the non-jetted population. As shown in \citet{jackson_exploring_2026}, the non-jetted and jet-dominated clusters meet at approximately the $\log R=2.5$ threshold; however this threshold may lead to contamination of the radio-quiet population by jet-dominated sources.\footnote{\citet{jackson_exploring_2026} aimed to maximise the purity of their jet-dominated sample and thus employed the GMM probabilities and $\log R = 2.5$ to remove sources likely not dominated by jets.} 
In the rest of the analysis, when attempting to isolate the radio-quiet quasars (i.e., those not dominated by jets), we remove sources with $\log R>2$. Two of the EHVOs fall into this category (1.9 per cent of the EHVOs radio-detected or otherwise) and 3.5 per cent of the parent sample (748 quasars). If instead we employ the Gaussian mixture model or $\log R=2.5$ cut to remove jets, the results remain in qualitative agreement with those presented here.

For clarity, throughout the paper we will refer to quasars with $\log R>2$ as `radio-loud', and these are quasars with radio emission likely dominated by jets. Quasars with detected radio emission but $\log R<2$ are sources that are `radio-quiet'; the origin of this radio emission is unclear. 

\section{Results}
\label{sec:results}
\subsection{Radio-detection fraction}
\subsubsection{Overall radio-detection fraction}

To assess whether the EHVOs have unique radio properties compared to quasars without EHVOs, we first compare the radio-detection fraction of the EHVO sample to the parent sample. As a whole, the EHVO sample has a $37.7^{+4.9}_{-4.4}$ per cent radio-detection fraction while only $24.0\pm0.3$ per cent of the parent sample is detected by LOFAR, a 1.57 factor increase in the radio-detection fraction of the EHVOs compared to the parent sample. For comparison \citet{morabito_origin_2019} found with LoTSS DR1 a 1.75 factor increase in the HiBALs over the non-BALs (3.33 factor for LoBALs), \citet{petley_connecting_2022} using LoTSS DR2 reported a 1.6 factor increase for HiBALs (2.22 for LoBALs), and Grieve et al. (in prep) using LoTSS DR3 a 1.18 factor increase for HiBALs. We additionally calculate the radio-detection fractions of the samples once radio-loud sources are removed: $36.5^{+6.9}_{-2.6}$ per cent of the EHVOs and $21.2\pm0.3$ per cent of the parent sample are radio-detected but radio-quiet.

\subsubsection{Radio-detection fraction for matched parent samples}

\begin{figure}
    \centering
    \includegraphics[width=\linewidth]{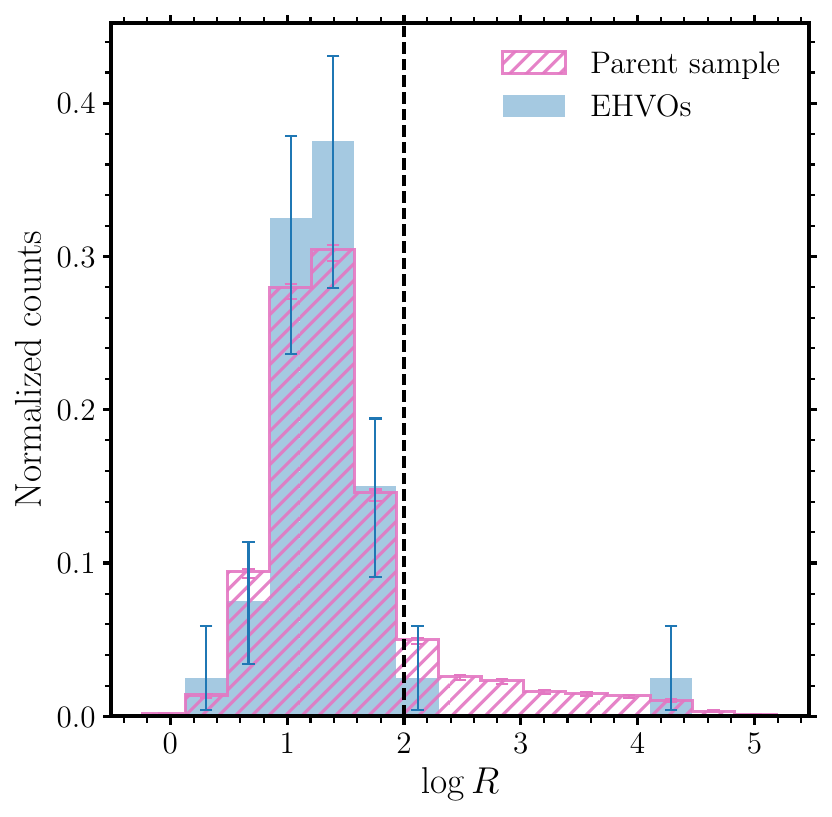}
    \caption{\change{$\log R$ distribution of the radio-detected EHVO (blue, filled) and parent (pink, hatched) samples. The errorbars are Poisson uncertainties following Gehrels' method for low number statistics \citep{Gehrels86_confidence_limits}. The dashed vertical line demarcates the radio-loud threshold. In the radio-quiet regime, the EHVO $\log R$ distribution is similar to that of the parent sample, i.e., the EHVOs exhibit radio luminosities that scale with their $L_{3000}$ in the same proportion as in the parent sample.}}
    \label{fig:logR}
\end{figure}

\change{As observed in Fig.~\ref{fig:lums}, while overlapping with the parent sample the EHVOs have higher $L_{3000}$ on average. As such, and given the correlation between optical luminosity and radio luminosity, the EHVOs may be more likely to be radio-detected purely due to being optically brighter. Indeed, the distribution of $\log R$ for the radio-detected EHVO sample is similar to that of the radio-detected parent sample (see Fig.~\ref{fig:logR}), implying that controlling for $L_{3000}$ may be enough to reproduce the high radio-detection fraction of the EHVOs. To examine the effect of $L_{3000}$ further, we produce subsamples of the parent sample matched to the EHVOs in $L_{3000}$ using \textsc{scikit-learn}'s nearest-neighbour algorithm \citep{scikit-learn}, keeping the 10 nearest non-EHVOs per EHVO. We confirm the success of the matching process via a two-sample Kolmogorov-Smirnov (KS) test ($p\text{-value} \gtrsim 0.99$)\footnote{In one dimension, we implement \textsc{scipy}'s two-sample KS test.} The radio-detection fraction of the matched parent sample is $29.1\pm1.4$ per cent which, while greater than the whole parent sample, is not within 1$\sigma$ of the EHVO radio-detection fraction, with the caveat that the EHVO uncertainties are large. Additionally, radio-loudness distribution of this $L_{3000}$-matched sample reveals a tentative shift to lower $\log R$ values implying that despite selecting quasars from the parent sample that are optically brighter than the average, their radio luminosities are brighter than the parent sample average but not as bright as the EHVOs (see Appendix~\ref{app:logR}). We postulate that some other additional variable is partly responsible for the EHVOs' increased radio-detection fraction.}

EHVOs occupy a particular volume of {\BHM}, {\lambdaEdd}, and {\CIV} emission parameter space and it is known that the radio properties are also correlated with these parameters. EHVOs are more likely to be found in objects with larger {\CIV} emission blueshifts \citepalias{hidalgo_connection_2022}, and, at \change{a} given {\BHM}, EHVOs lie in the upper half of the distribution of {\lambdaEdd} of the parent sample (Flores et al. in prep). As such, EHVOs lie in regions of {\BHM}, {\lambdaEdd}, and {\CIV} blueshift parameter space that are known to have a higher radio detection rate (\citetalias{rankine_placing_2021}; \citealt{Richards21_probing_wind, jackson_exploring_2026}). The question is, therefore, can the high radio-detection fraction in the EHVOs be explained by any of these parameters?

\begin{figure}
    \centering
    \includegraphics[width=\linewidth]{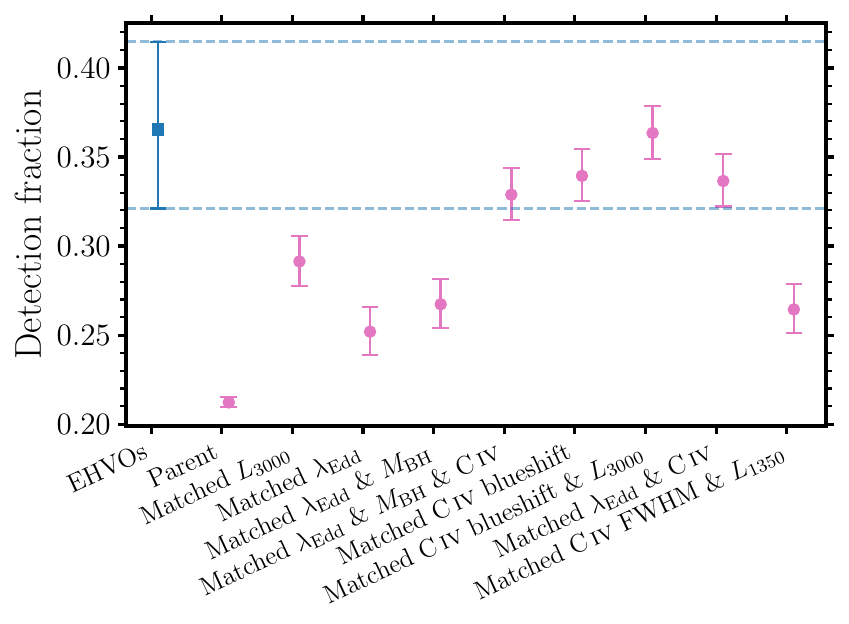}
    \caption{\change{Graphical version of Table~\ref{tab:rdf}: radio-detection fraction for the EHVO sample, the parent sample, and each of the matched parent samples. Likely-jetted sources are removed by excluding those with $\log R >2$. For brevity, `{\CIV} blueshift' is sometimes shortened to `{\CIV}'. Matching in {\CIV} blueshift is sufficient to produce a sample with a similar radio-detection fraction to that of the EHVO sample.}
    }
    \label{fig:rdf}
\end{figure}

\change{To answer this question, we produce \change{subsamples of the parent sample} matched to the EHVOs in {\BHM}, {\lambdaEdd}, and {\CIV} emission blueshift. The KS tests all give $p\text{-values} \gtrsim 0.99$, confirming success of the matching.\footnote{In two dimensions, we implement \textsc{ndtest} (written by Zhaozhou Li, \url{https://github.com/syrte/ndtest}) and \textsc{fasano-franceschini-test} in three dimensions \citep[][\url{https://github.com/wmpg/fasano-franceschini-test}]{chow_decameter-sized_2025} which are based on \citet{peacock_two-dimensional_1983} and \citet{fasano_multidimensional_1987}.}}
Table~\ref{tab:rdf} presents the radio-detection fractions along with $1\sigma$ binomial uncertainties \citep[detailed in][]{Cameron11_estimation_confidence} for the EHVOs, full parent sample, and parent samples matched on a combination of parameters. Figure~\ref{fig:rdf} presents a graphical version of these results. \change{Matching in {\CIV} blueshift alone is sufficient to recreate the radio-detection fraction of the whole EHVO sample; however, matching jointly in {\CIV} blueshift and $L_{3000}$ produces the radio-detection fraction in most agreement with that of the EHVOs.}

The radio-detection fraction of the parent samples matched in {\CIV} blueshift and {\lambdaEdd} together, and additionally with {\BHM} are not significantly different from that of the sample matched only in {\CIV} blueshift. 
\change{A correlation exists between {\CIV} blueshift and {\lambdaEdd} therefore it is possible that by matching on {\CIV} blueshift we are preferentially selecting truly high-{\lambdaEdd} sources. Given the uncertainty on single-epoch black hole mass measurements, the lack of a consistent match in radio-detection fraction when matching only on {\lambdaEdd} does not constitute evidence that {\lambdaEdd} is unrelated to the high radio-detection fraction in the EHVOs.}

Since {\lambdaEdd} and {\BHM} are inherently correlated, we also match on the parameters that are used to estimate these quantities, namely the {\CIV} FWHM and $L_{1350}$. The resulting detection fraction is consistent with that when matched on {\lambdaEdd} and {\BHM}. 

\begin{table}
  \caption{\change{Radio-detection fraction of the parent sample, EHVOs, and the parent sample matched in $L_{3000}$, {\BHM}, {\lambdaEdd}, and {\CIV} blueshift. The distribution of $L_{3000}$ in the EHVO population contributes to the observed high fraction of radio detections, as does the {\CIV} blueshift. Matching in $L_{3000}$ and {\CIV} blueshift together provides the most accurate reproduction of the EHVO radio-detection fraction. Quoted uncertainties are the 1$\sigma$ binomial uncertainties, derived from a Beta($N_\text{det}+1$, $N_\text{tot}-N_\text{det}+1$) posterior with a uniform prior \citep[as recommended by][]{Cameron11_estimation_confidence}.}}
    \centering
    \begin{tabular}{rcc}
\hline
 & $N_\text{det}/N_\text{tot}$ & \begin{minipage}{0.17\columnwidth}\centering Detection fraction\end{minipage}\\
\hline
EHVOs & 38/104 & \change{$0.37_{-0.04}^{+0.05}$}\\
Parent & 4360/20545 & $0.21_{-0.00}^{+0.00}$\\
\change{Matched $L_{3000}$} & 303/1040 & $0.29_{-0.01}^{+0.01}$\\
Matched {\lambdaEdd} & 262/1040 & $0.25_{-0.01}^{+0.01}$\\
Matched {\BHM} \& {\lambdaEdd} & 278/1040 & $0.27_{-0.01}^{+0.01}$\\
Matched {\BHM} \& {\lambdaEdd} \& \change{{\CIV} blueshift} & 342/1040 & $0.33_{-0.01}^{+0.01}$\\
Matched {\CIV} blueshift & 353/1040 & $0.34_{-0.01}^{+0.01}$\\
\change{Matched {\CIV} blueshift \& $L_{3000}$} & 378/1040 & $0.36_{-0.01}^{+0.02}$\\
Matched {\lambdaEdd} \& {\CIV} \change{blueshift} & 350/1040 & $0.34_{-0.01}^{+0.01}$\\
Matched {\CIV} FWHM \& $L_{1350}$ & 275/1040 & $0.26_{-0.01}^{+0.01}$\\
\hline
\end{tabular}
    \label{tab:rdf}
\end{table}

\subsubsection{Radio-detection fraction as a function of {\CIV} blueshift}

\begin{figure}
    \centering
    \includegraphics[width=\linewidth]{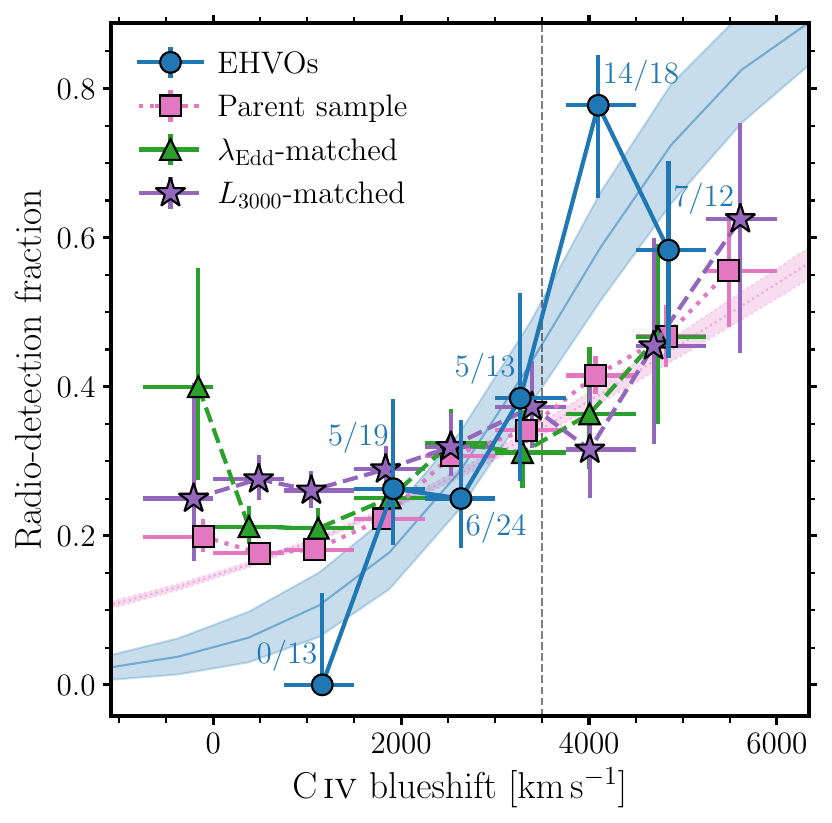}
    \caption{Radio-detection fraction in bins of {\CIV} blueshift for the EHVOs (blue circles and numbers denoting the fraction), the parent sample (pink squares), and the sample matched in {\lambdaEdd} (green triangles), \change{and $L_{3000}$} (purple stars). We plot only the bins containing $\geq$5 objects. Horizontal error bars are the width of the {\CIV} blueshift bins and the points are placed at the median {\CIV} blueshift. Vertical error bars are the 1$\sigma$ binomial uncertainties following \citet{Cameron11_estimation_confidence}. The grey dashed vertical line marks the maximum {\CIV} blueshift explored in \citetalias{rankine_placing_2021}. \change{The shaded curves show the results of the linear logistic fit: mean and $1\sigma$ uncertainties estimated from 2000 bootstrap samples from the (assumed) multivariate normal parameter covariance matrix.}}
    \label{fig:civ_det}
\end{figure}

In Fig.~\ref{fig:civ_det}, we plot the radio-detection fraction in bins of {\CIV} blueshift (750\,km\,s$^{-1}$-wide) for the EHVOs, parent sample, and a select few matched samples. Note, we plot only bins with $\geq$5 objects.
In general, the radio-detection fraction increases as {\CIV} blueshift increases in all populations, but the EHVO sample is markedly different from the parent sample, matched or otherwise. Specifically, below {\CIV} blueshifts of $\sim$2000\,km\,s$^{-1}$, the EHVO radio-detection fraction is lower than the whole or matched parent sample, and the opposite is true at {\CIV} blueshifts between 3500--5000\,km\,s$^{-1}$. There is also a drop in the radio-detection fraction at the largest {\CIV} blueshifts.
Matching in the various parameters does not significantly change the radio-detection fraction in a given {\CIV} blueshift bin. However, the global detection fraction does change (Fig.~\ref{fig:rdf} and Table~\ref{tab:rdf}) due to the change in the relative contribution of each {\CIV} bin to the global average. The parent sample is weighted towards low blueshifts while there are more EHVOs at higher blueshift and these populations dominate the overall radio-detection fractions in their respective samples. 

\change{
To assess quantitatively whether the radio-detection fraction as a function of {\CIV} blueshift differs between the EHVO and parent samples, we fit a logistic regression model including an overall blueshift term and an EHVO-specific blueshift interaction term. Specifically, we fit the model
\begin{equation}
    \begin{aligned}
        \text{logit}\left[P(Y_{i}=1)\right] = \beta_0 + \beta_1\left(\frac{\text{{\CIV} blueshift}_i}{10^3\,\text{km}\,\text{s}^{-1}}\right) + \beta_2\text{EHVO}_i \\ 
        + \beta_3\left(\frac{\text{{\CIV} blueshift}_i}{10^3\,\text{km}\,\text{s}^{-1}}\right)\text{EHVO}_i,
    \end{aligned}
\end{equation}
where $Y_i=1$ indicates a radio detection and $\text{EHVO}_i=1$ if object $i$ is an EHVO and 0 otherwise. $\beta_0$ and $\beta_1$ are the intercept and slope in the non-EHVO population. $\beta_2$ is the offset for the EHVO population and $\beta_3$ is the EHVO-specific interaction term such that the model for the EHVO subsample has intercept $\beta_0 + \beta_2$ and slope $\beta_1 + \beta_3$. Table~\ref{tab:logit} contains the fitted parameters and their significance for the EHVO+parent sample. $\beta_2$ and $\beta_3$ are significantly non-zero, suggesting that the EHVO population exhibits a steeper increase in radio-detection fraction with {\CIV} blueshift than the parent population. The resulting fitted models for the EHVO and parent samples are shown as the shaded curves encompassing the $1\sigma$ uncertainties estimated from 2000 bootstrap samples assuming a multivariate normal distribution described by the covariance matrix of the $\beta$ parameters.}

\begin{table}
    \caption{\change{Fitted parameters and associated $p$-values from the logistic regression model (Equation 2), for the EHVO+parent sample. $\beta_0$ and $\beta_1$ are the intercept and blueshift slope for the non-EHVO population; $\beta_2$ is the offset for the EHVO population; and $\beta_3$ is the EHVO-specific blueshift interaction term. Estimates are the mean with $1\sigma$ uncertainties. The $p$-values correspond to two-tailed tests of the null hypothesis that each parameter is zero. Significantly non-zero values of $\beta_2$ and $\beta_3$ indicate that the EHVO population has both a different normalisation and a steeper dependence of radio-detection fraction on {\CIV} blueshift compared with the parent sample.}}
    \label{tab:logit}
    \centering
    \begin{tabular}{ccc}
\hline
Parameter & Estimate $\pm1\sigma$ & $p$-value \\
\hline
$\beta_0$ & $-1.770\pm0.031$ & $<0.001$ \\
$\beta_1$ & $0.321\pm0.017$ & $<0.001$ \\
$\beta_2$ & $-1.415\pm0.686$ & $0.039$ \\
$\beta_3$ & $0.539\pm0.204$ & $0.008$ \\
\hline
\end{tabular}    
\end{table}

\change{The logistic model above has the advantage of being fitted on the individual unbinned data; however, assumes linearity, which may not accurately capture the true shape of the relationship. We additionally perform a model-independent stratified permutation test, by shuffling the EHVO labels within each {\CIV} blueshift bin, to test the null hypothesis that EHVO status does not affect radio detection within a given blueshift bin (i.e., that EHVO labels are exchangeable). For the observed data, we calculate the test statistic,
\begin{equation}
    T=\sum_b N_{\text{EHVO},b}
\left(\text{RDF}_{\text{EHVO},b}-\text{RDF}_{\text{parent},b}\right)^2,
\end{equation}
where $N_{\text{EHVO},b}$ is the number of EHVOs in {\CIV} blueshift bin $b$ and $\text{RDF}_{\text{EHVO},b}$ and $\text{RDF}_{\text{parent},b}$ are the radio-detection fractions in bin $b$ for the EHVO and parent samples, respectively. We then produce 5000 permutations of the EHVO labels within each bin and compute $T_\text{perm}$ as above for each permutation. The observed test statistic is $T_\text{obs} = 3.32$ and the distribution of $T_\text{perm}$ has mean 1.33 and standard deviation 0.77. We calculate the associated $p$-value as follows:
\begin{equation}
    p=\frac{N(T_{\text{perm}}\geq T_{\text{obs}})+1}
{N_{\text{perm}}+1}.
\end{equation}
The resulting $p\text{-value}$ is 0.0218 suggesting we can reject, at the 5 per cent significance level, the null hypothesis that the EHVO radio-detection fraction as a function of {\CIV} blueshift is statistically indistinguishable from that of the parent population.
}

In summary, there are persistent, tentatively significant differences in the radio-detection fraction between the EHVO and parent samples. This shows that, while the \textit{global} radio-detection fraction can be recreated, controlling for {\CIV} blueshift in the parent population is \change{potentially} insufficient to explain all of the differences in radio-detection fraction of the EHVOs; \change{however, we note the small sample size of the EHVO population.}

\subsection{Correlations between EHVO absorption properties and radio}

\begin{figure}
    \centering
    \includegraphics[width=\linewidth]{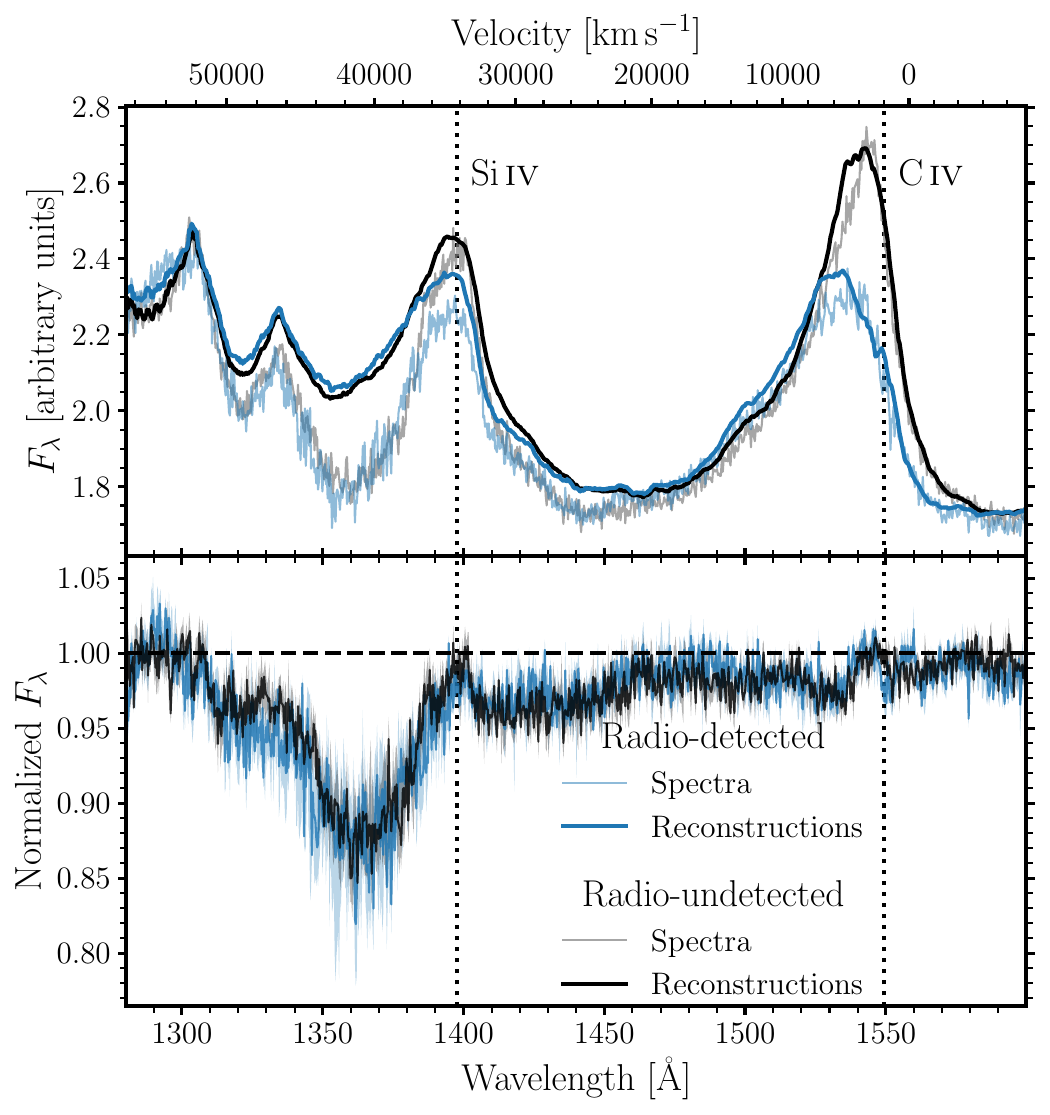}
    \caption{Median composite spectra of the radio-detected (blue) and undetected EHVOs (black). Top: the median spectra (light blue and grey) and median reconstructions (dark blue and black).
    Bottom: the median composite of the spectra normalized by the reconstructions. The EHVO troughs are very similar between the radio-detected and undetected composites. However, the {\CIV} emission lines differ: the radio-detected quasars have, on average, larger {\CIV} blueshifts, in agreement with previous studies.}
    \label{fig:comps}
\end{figure}

\citet{petley_connecting_2022} found differences in the BAL absorption profiles between the median composites of radio-detected and undetected quasars, specifically there is a shift of the deepest part of the trough to bluer wavelengths (higher speeds) and more absorption is present at bluer wavelengths generally -- resulting in a broader trough -- in the radio-detected composite. These differences are more pronounced in the LoBAL composites.  
However, Grieve et al. (in prep) show that binning by {\lambdaEdd} and {\BHM} reduces much of the differences observed in the HiBALs (LoBALs are not studied). The emission profiles are also very similar when binned in this way.
While we do not have the sample sizes to bin by any property, we make median composites of the whole sample of radio-detected and undetected EHVO spectra, reconstructions, and, separately, the spectra normalized by their individual reconstructions. The 1300--1600\,{\AA} region of these composites are shown in Fig.~\ref{fig:comps}. As expected, the {\CIV} \textit{emission} is more highly blueshifted and weaker in the radio-detected composite.
However, the radio-detected and radio-undetected EHVOs do not differ in their absorption profiles, in agreement with the BAL results of Grieve et al. (in prep).
In contrast to Grieve et al. (in prep), binning in {\lambdaEdd} and {\BHM} is not required to produce similar radio-detected and undetected absorption profiles due to the restricted range in this parameter space that EHVOs are found.

\begin{figure}
    \centering
    \includegraphics[width=\linewidth]{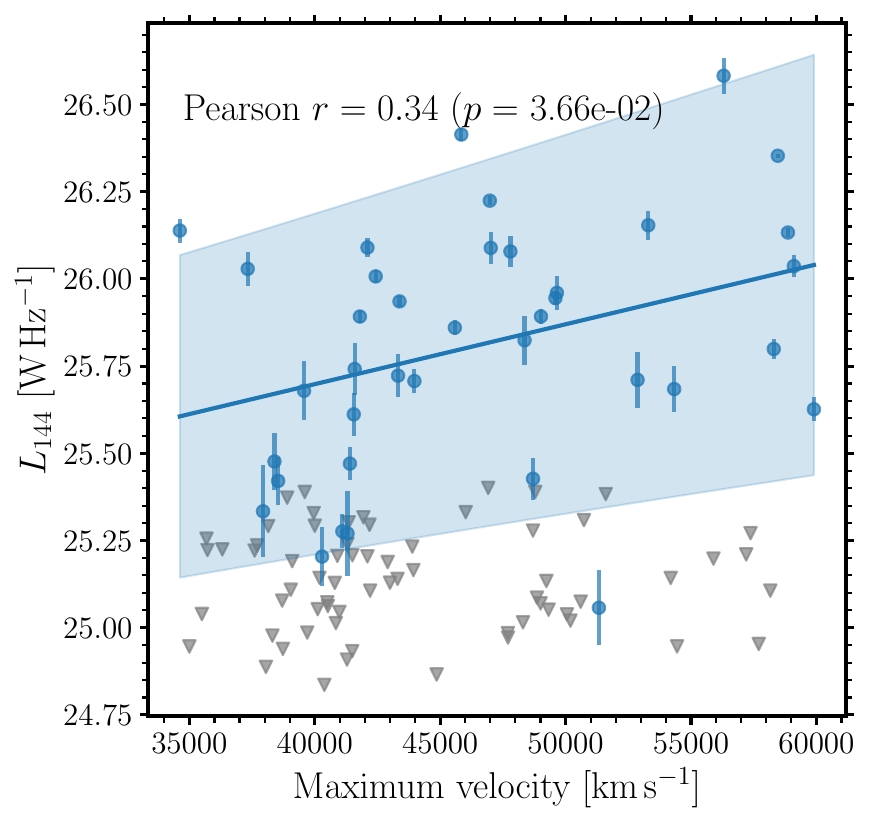}
    \caption{\change{Radio luminosity as a function of maximum absorption velocity. Upper limits for the radio-undetected EHVOs are plotted as triangles. The best-fitting linear regression model to the detected radio-quiet sample is shown, and the shaded region represent the corresponding $1\sigma$ uncertainties.
    } 
    }
    \label{fig:lum_ehvo}
\end{figure}

Despite the lack of any noticeable differences between the radio-detected and undetected composite EHVO absorption, there is a \change{tentative} correlation between the radio luminosity and the maximum absorption velocity of the radio-detected subsample (Pearson correlation coefficient $r=0.34$, $p\text{-value}=3.66\text{e-2}$), as shown in Fig.~\ref{fig:lum_ehvo}. A statistically insignificant correlation with the balnicity index of the EHVOs exists ($r=0.22$, $p\text{-value}=1.82\text{e-1}$). Upper limits on the radio luminosity for the undetected sample are estimated assuming a 0.35\,mJy flux limit. These limits suggest that there is significant scatter in radio luminosity across absorption velocity.

\section{Discussion}
\label{sec:disc}

EHVOs are still a mystery, as is the origin of radio-quiet emission in quasars. Previous results have shown that the radio-detection fraction increases with increasing {\CIV} blueshift and, separately, that EHVOs have large {\CIV} blueshifts. We thus wanted to investigate if the EHVOs have special radio properties compared to the general quasar population, and, if so, whether this can be explained by blueshift and/or {\lambdaEdd}.

In summary of our results, the radio-detection fraction in the EHVO population as a whole is greater than that of the parent sample, unless {\CIV} blueshift is controlled (see Fig.~\ref{fig:rdf}). However, there are still remaining differences between the EHVO radio-detection fraction and the parent sample as a function of {\CIV} emission line blueshift (see Fig.~\ref{fig:civ_det}). The absorption profiles in the radio-detected EHVOs do not show significant differences from the radio-undetected EHVOs; however, their {\CIV} emission is more blueshifted (Fig.~\ref{fig:comps}).
Finally, we find a \change{tentative} correlation between radio luminosity and the EHVO maximum outflow speed (Fig.~\ref{fig:lum_ehvo}).

In Section~\ref{sec:disc:orig} we discuss the possible origin of the radio emission in the EHVO sample and argue for a wind origin due to the trend with {\CIV} blueshift. We then consider our results under this wind-origin lens in Section~\ref{sec:disc:ehvo}.

\subsection{Previous work on the origin of the radio emission}
\label{sec:disc:orig}

Weak/compact jets, star formation, and quasar winds are all plausible mechanisms for the production of the radio emission observed in the quasars in our sample.
For a similar parent sample, \citetalias{rankine_placing_2021} looked at the possible contribution of all three finding that all are capable of producing the observed radio emission (see their section 4.1 and relevant references therein). Here, we consider their points in the context of our sample and suggest that all three are likely playing some part as well.

\textbf{Winds. }\citetalias{rankine_placing_2021} show that the wind-shock model of \citet{Nims15_observational_signatures} produces a correlation between radio luminosity and bolometric luminosity that is consistent with that observed in the radio-quiet quasars.
Winds seem to be the simplest explanation for the radio emission across all of our samples due to the correlation with {\CIV} blueshift. 

\textbf{Jets. }It is possible that weak and/or compact jets contaminate our radio-quiet population and \citet{jackson_exploring_2026} have shown that strong jets at least are correlated with {\BHM} and {\lambdaEdd} (and anti-correlated with {\CIV} blueshift). Under the magnetically-dominated accretion disc model of \citet{hopkins_multi-phase_2024}, the magnetic flux threading the disc is responsible for launching jets stochastically and intermittently, but with some correlation with the accretion rate and/or {\BHM}. Accretion rate is correlated with {\CIV} blueshift, and so perhaps this correlation between blueshift and radio-detection fraction is a result of the jet--accretion rate correlation.

\textbf{Star formation. }Through simple Monte Carlo simulations, \citetalias{rankine_placing_2021} show that the radio emission in the radio-quiet quasars could be produced from star formation alone if median SFRs of $\sim$30\,M\,yr$^{-1}$ in all quasars in their sample, and $\sim$300--1000\,M\,yr$^{-1}$ in the radio-detected sources, can be accommodated. The increasing radio-detection fraction with increasing {\CIV} blueshift could be driven by the cold gas supply to nuclear regions. An abundance of available cold gas would increase the star formation rate, but also make available more fuel for the quasar. 

In either scenario where jets or star formation is the dominant producer of radio emission in the radio-quiet quasars, and jets or star formation are correlated with {\BHM}, {\lambdaEdd}, and (anti-)correlated with {\CIV} blueshift, then it is possible that when controlling for these parameters we are merely controlling for the types of jets or star formation rate.
However, the effect of matching on {\CIV} blueshift alone would suggest a tight relationship with star formation -- to produce the statistically similar match in radio-detection fraction -- that seems implausible given the variety of timescales and physical scales at play.

\subsection{Can the radio emission be directly linked to the EHVO?}
\label{sec:disc:ehvo}

EHVOs exhibit a higher radio-detection fraction when compared to their parent sample. We are able to account for most of this excess in the global radio-detection fraction when controlling for {\CIV} blueshift, {\lambdaEdd} and {\BHM} (in order of importance). Additionally, the radio-detected and undetected EHVO subsamples display no discernable differences in their EHVO absorption properties. A straightforward interpretation of these results would be that the underlying correlation between {\CIV} blueshift and radio-detection fraction is driving the increased detection fraction in the EHVOs as they typically have large blueshifts.
This could also suggest that the outflow in absorption is not responsible for the radio emission. 
This interpretation would be possible if we always observed EHVOs whenever there is a fast wind, but this is unlikely. \change{Instead, non-EHVO quasars -- particularly those with large {\CIV} emission blueshifts -- may have EHVOs but along other lines-of-sight. The similar radio-detection fractions (albeit with tentative differences) and $\log R$ distributions (see Appendix~\ref{app:logR}) suggest these are indeed the same objects, observed at different lines-of-sight.}

Monte Carlo Radiative Transfer simulations by \citet{dannen_wind_2026} show that EHVOs are observed over a narrow range of inclination viewing angles towards the quasar/accretion disc. Here, the authors discuss that EHVO production requires the gas to reach its terminal velocity over short distances (short acceleration lengths) while the gas remains confined, and this only occurs for a short range of viewing angles. This confined gas, moving at very high speeds, could be responsible for producing shocks that would result in radio emission.

However, \citet{PRH_2025} places the EHVO of J164653.72+243942.2 at $\sim$5--28 parsec from the black hole. It is possible that these close-in wind features do not survive to reach the kiloparsec-scales for collision with the ISM that would produce the radio emission. Alternatively, if the wind is clumpy \citep[e.g.,][]{filiz_ak_broad_2012, Matthews16_testing_quasar, Matthews26_how_massive, Vivek26_recurrent_multiyear}, shocks within the medium itself may produce radio emission: \citet{sotomayor_nonthermal_2022} show that, in super-accreting sources, synchrotron radiation can be produced in bow shocks that form around the BLR clouds. \change{However, emission from BLR-scale shocks is unlikely to explain the 144\,MHz detections: it must escape the dense ionized gas of the BLR, where free-free absorption makes the region opaque at these frequencies, and such compact regions would in any case be self-absorbed, with spectra peaking at GHz rather than MHz frequencies (\citealt{Kellermann81_compact_radio}; and fig. 4 of \citealt{Raginski16_agn_coronal} by analogy with compact coronal emission).} In either case, a weak correlation between the radio luminosity and EHVO velocity may be expected since the EHVO would not be tracing directly the kiloparsec-scale wind or the global BLR wind properties, and this is what we find (see Fig.~\ref{fig:lum_ehvo}).

More puzzling is the observed distribution of radio-detection fraction with {\CIV} emission blueshift (Fig.~\ref{fig:civ_det}). While this is the property that best matches the overall EHVO radio detection, we observe a difference between the distributions of EHVOs and the {\CIV} parent sample when binned by {\CIV} emission blueshift.
Whatever is the cause of the underlying correlation between {\CIV} blueshift and radio-detection fraction, the steeper slope with blueshift in the EHVOs hints at a wind origin for at least this remaining trend with {\CIV} blueshift.

\section{Conclusions and Future Work} 
\label{sec:concl}

In order to investigate quasar winds as a contributor to the radio emission, we have examined the LOFAR radio properties of the fast and energetic Extremely High Velocity Outflows, travelling at speeds 0.1--0.2$c$ and detected in the optical/UV spectra of luminous quasars. We have cross-matched the SDSS DR9 and DR16 EHVO sample from \citet{hidalgo_survey_2020} and Candelaria-Stoner et al. (in prep) with the DR3 of the LOFAR Two-metre Sky Survey \citep{shimwell_lofar_2026}. We summarise our key results:

\begin{enumerate}
    \item 21\,293 quasars in the parent sample and 106 EHVOs fall within the footprint of the LoTSS DR3 and have spectrum reconstructions following the approach in \citet{Rankine20_bal_non-bal}.
    \item 40 EHVOs (38 per cent) and 5108 (24 per cent) of the parent sample are detected in LoTSS at $5\sigma$, marking a 1.57 factor increase in the overall radio-detection fraction in the EHVOs compared to the parent sample.
    \item \change{The radio-detected EHVO and parent samples have similar radio-loudness distributions ($\log R$; Fig.~\ref{fig:logR}) such that matching on $L_{3000}$ increases the radio-detection fraction of the parent sample substantially; however, not sufficiently to explain all of the increased radio-detection fraction of the EHVOs.}
    \item Matching the parent sample to the EHVOs with {\CIV} blueshift is sufficient to match the \textit{global} EHVO radio-detection fraction while matching in {\lambdaEdd} is not (see Fig.~\ref{fig:rdf}). However, differences still remain as a function of {\CIV} blueshift (see Fig.~\ref{fig:civ_det}), where EHVOs show a steeper distribution.
    \item Median composite EHVO spectra reveal no significant differences between the profiles of the EHVOs detected and undetected in radio (see Fig.~\ref{fig:comps}). However, there is a \change{tentative} correlation between the maximum absorption velocity and the radio luminosity (see Fig.~\ref{fig:lum_ehvo}).
\end{enumerate}

If winds account for a component of the radio emission from quasars, then these findings suggest that the EHVOs correspond to a subset of quasars with fast winds, viewed such that our line-of-sight passes through the outflowing material. 
Future work on cases at even larger speeds, such as the EHVOs found in PDS 456 \citep{hamann_2018} and SDSS J231854.31+243954.2 \citep{seaton_new_2026}, will allow us to expand this even further.\footnote{PDS 456 is not in the LoTSS DR3 footprint and  J2318 does not have a radio counterpart in LoTSS DR3.}

Throughout this paper we have assumed a radio spectral index of -0.7 which is what is expected for \change{optically-thin} synchrotron emission. \change{In their VLA-observed sample of BAL quasars, \citet{Chen26_radio_emission} find five to have steep spectral slopes at 5.5--9\,GHz which they show could be associated with the BAL outflow. Another five have flat or inverted radio spectra which may be consistent with intrinsically steep slopes flattened by free-free absorption in the outflow. Radio SEDs of EHVO quasars have not been studied in detail to date.} Constraining the spectral indices of the EHVOs with multi-frequency radio observations would help to address the origin of their radio emission.

Additionally, the EHVOs represent an interesting sample to examine with the sub-arcsec resolution International \change{LOFAR} Telescope \citep{Morabito25_decade_sub-arcsecond} to further distinguish the radio emission from small-scale jets, star-formation and/or quasar winds.

\change{If wind-driven shocks are contributing to the radio emission, then questions still remain about how or if the wind traced by the EHVO extends to scales larger than the BLR where this low-frequency radio emission may originate. Studying the narrow line region (NLR) of quasars hosting EHVOs will help to assess the impact of the EHVOs. Such studies of BAL quasars have been performed with results tentatively suggesting that the [\ion{O}{iii}] emission and {\CIV} BALs are tracing different parts of the same wind \citep[e.g.,][]{Xu20_evidence_that, Temple24_o_iii}. Meanwhile \citet{Escott25_unveiling_agn, Escott26_sub-arcsecond_ilt} have revealed correlations between the LOFAR radio data (6-arcsec and sub-arcsec resolution) and [\ion{O}{iii}] outflow properties. Overall, these results motivate connecting EHVOs to NLR-scale outflows to clarify the origin of the low-frequency radio emission in the EHVO population.}

\section*{Acknowledgements}

We thank Gordon Richards and Gregory Walsh for very helpful discussions, and Kenneth Duncan and James Petley for early guidance. \change{We thank the anonymous referee for their constructive review of the paper and suggestions that helped to improve the paper.}

ALR acknowledges support from a Leverhulme Early Career Fellowship.
PRH acknowledges support from the National Science Foundation and the Sloan Digital Sky Survey's FAST 4ward program through award AST-2425222.

LOFAR data products were provided by the LOFAR Surveys Key Science project (LSKSP; \url{https://lofar-surveys.org/}) and were derived from observations with the International LOFAR Telescope (ILT). LOFAR (van Haarlem et al. 2013) is the Low Frequency Array, designed and constructed by ASTRON. It has observing, data processing, and data storage facilities in several countries, which are owned by various parties (each with their own funding sources), and which are collectively operated by the LOFAR ERIC under a joint scientific policy. The efforts of the LSKSP have benefited from funding from the European Research Council, NOVA, NWO, CNRS-INSU, the SURF Co-operative, the UK Science and Technology Funding Council and the Jülich Supercomputing Centre.

Funding for the Sloan Digital Sky 
Survey IV has been provided by the 
Alfred P. Sloan Foundation, the U.S. 
Department of Energy Office of 
Science, and the Participating 
Institutions. 

SDSS-IV acknowledges support and 
resources from the Center for High 
Performance Computing  at the 
University of Utah. The SDSS 
website is www.sdss4.org.

SDSS-IV is managed by the 
Astrophysical Research Consortium 
for the Participating Institutions 
of the SDSS Collaboration including 
the Brazilian Participation Group, 
the Carnegie Institution for Science, 
Carnegie Mellon University, Center for 
Astrophysics | Harvard \& 
Smithsonian, the Chilean Participation 
Group, the French Participation Group, 
Instituto de Astrof\'isica de 
Canarias, The Johns Hopkins 
University, Kavli Institute for the 
Physics and Mathematics of the 
Universe (IPMU) / University of 
Tokyo, the Korean Participation Group, 
Lawrence Berkeley National Laboratory, 
Leibniz Institut f\"ur Astrophysik 
Potsdam (AIP),  Max-Planck-Institut 
f\"ur Astronomie (MPIA Heidelberg), 
Max-Planck-Institut f\"ur 
Astrophysik (MPA Garching), 
Max-Planck-Institut f\"ur 
Extraterrestrische Physik (MPE), 
National Astronomical Observatories of 
China, New Mexico State University, 
New York University, University of 
Notre Dame, Observat\'ario 
Nacional / MCTI, The Ohio State 
University, Pennsylvania State 
University, Shanghai 
Astronomical Observatory, United 
Kingdom Participation Group, 
Universidad Nacional Aut\'onoma 
de M\'exico, University of Arizona, 
University of Colorado Boulder, 
University of Oxford, University of 
Portsmouth, University of Utah, 
University of Virginia, University 
of Washington, University of 
Wisconsin, Vanderbilt University, 
and Yale University.

%%%%%%%%%%%%%%%%%%%%%%%%%%%%%%%%%%%%%%%%%%%%%%%%%%
\section*{Data Availability}
 
The EHVO parent samples and the EHVO properties are or will be made available through \citet{hidalgo_survey_2020} and Candelaria-Stoner et al. (in prep). The LoTSS DR3 catalogue is available at \url{https://lofar-surveys.org/}. Measurements made from the ICA-based spectrum reconstructions will be made available on request to the corresponding author.



\bibliographystyle{mnras}
\bibliography{references} % if your bibtex file is called example.bib

@string{june = {June}}

@article{Baldwin96_very_high,
 adsurl = {https://ui.adsabs.harvard.edu/abs/1996ApJ...461..664B},
 author = {{Baldwin}, J.~A. and {Ferland}, G.~J. and {Korista}, K.~T. and {Carswell}, R.~F. and {Hamann}, F. and {Phillips}, M.~M. and {Verner}, D. and {Wilkes}, Belinda J. and {Williams}, R.~E.},
 doi = {10.1086/177093},
 journal = {\apj},
 month = {April},
 pages = {664},
 title = {{Very High Density Clumps and Outflowing Winds in QSO Broad-Line Regions}},
 volume = {461},
 year = {1996}
}

@software{baumann_2026_mocpy,
 author = {Baumann, Matthieu and
Boch, Thomas and
Marchand, Manon and
Pineau, François-Xavier},
 month = {March},
 publisher = {Zenodo},
 swhid = {swh:1:dir:3c1001af7499fd6c8a03265b6e6f03f0789f6bc0
;origin=https://doi.org/10.5281/zenodo.7637180;vis
it=swh:1:snp:a5fbab370f33cb0e7c48549d21db4c398abdd
845;anchor=swh:1:rel:ade67bb63f22b2124ae673b2a3f04
c2638b9bc6b;path=cds-astro-mocpy-ca0a04a},
 title = {MOCPy},
 url = {https://doi.org/10.5281/zenodo.18935153},
 version = {v0.20.0},
 year = {2026}
}

@article{Becker95_first_survey,
 adsurl = {https://ui.adsabs.harvard.edu/abs/1995ApJ...450..559B},
 author = {{Becker}, Robert H. and {White}, Richard L. and {Helfand}, David J.},
 doi = {10.1086/176166},
 journal = {\apj},
 month = {September},
 pages = {559},
 title = {{The FIRST Survey: Faint Images of the Radio Sky at Twenty Centimeters}},
 volume = {450},
 year = {1995}
}

@article{Bicknell02_connections_between,
 adsurl = {https://ui.adsabs.harvard.edu/abs/2002NewAR..46..365B},
 author = {{Bicknell}, Geoffrey V.},
 doi = {10.1016/S1387-6473(01)00210-X},
 journal = {\nar},
 month = {May},
 number = {2-7},
 pages = {365-379},
 title = {{Connections between jet physics and the properties of radio-loud and radio-quiet galaxies}},
 volume = {46},
 year = {2002}
}

@article{Cameron11_estimation_confidence,
 adsurl = {https://ui.adsabs.harvard.edu/abs/2011PASA...28..128C},
 archiveprefix = {arXiv},
 author = {{Cameron}, Ewan},
 doi = {10.1071/AS10046},
 eprint = {1012.0566},
 journal = {\pasa},
 month = {June},
 number = {2},
 pages = {128-139},
 primaryclass = {astro-ph.IM},
 title = {{On the Estimation of Confidence Intervals for Binomial Population Proportions in Astronomy: The Simplicity and Superiority of the Bayesian Approach}},
 volume = {28},
 year = {2011}
}

@article{chow_decameter-sized_2025,
 adsurl = {https://ui.adsabs.harvard.edu/abs/2025Icar..42916444C},
 archiveprefix = {arXiv},
 author = {{Chow}, Ian and {Brown}, Peter G.},
 doi = {10.1016/j.icarus.2024.116444},
 eid = {116444},
 eprint = {2501.03308},
 journal = {\icarus},
 month = {March},
 pages = {116444},
 primaryclass = {astro-ph.EP},
 title = {{Decameter-sized Earth impactors ─ I: Orbital properties}},
 volume = {429},
 year = {2025}
}

@article{coatman_correcting_2017,
 adsurl = {https://ui.adsabs.harvard.edu/abs/2017MNRAS.465.2120C},
 archiveprefix = {arXiv},
 author = {{Coatman}, Liam and {Hewett}, Paul C. and {Banerji}, Manda and {Richards}, Gordon T. and {Hennawi}, Joseph F. and {Prochaska}, J. Xavier},
 doi = {10.1093/mnras/stw2797},
 eprint = {1610.08977},
 journal = {\mnras},
 month = {February},
 number = {2},
 pages = {2120-2142},
 primaryclass = {astro-ph.GA},
 title = {{Correcting C IV-based virial black hole masses}},
 volume = {465},
 year = {2017}
}

@article{Condon92_radio_emission,
 adsurl = {https://ui.adsabs.harvard.edu/abs/1992ARA&A..30..575C},
 author = {{Condon}, J.~J.},
 doi = {10.1146/annurev.aa.30.090192.003043},
 journal = {\araa},
 month = {January},
 pages = {575-611},
 title = {{Radio emission from normal galaxies.}},
 volume = {30},
 year = {1992}
}

@article{dannen_wind_2026,
 adsurl = {https://ui.adsabs.harvard.edu/abs/2026ApJ..1007L..31D},
 archiveprefix = {arXiv},
 author = {{Dannen}, Randall C. and {Proga}, Daniel and {Hidalgo}, Paola Rodr{\'\i}guez and {Smith}, Kara and {Ritchie}, Anna and {Flores}, Liliana and {Kalet}, Morrigan},
 doi = {10.3847/2041-8213/ae9079},
 eid = {L31},
 eprint = {2606.26305},
 journal = {\apjl},
 month = {August},
 number = {2},
 pages = {L31},
 primaryclass = {astro-ph.GA},
 title = {{Wind Acceleration as a Driver of Detached Blueshifted Absorption in Quasar Disk Winds}},
 volume = {1007},
 year = {2026}
}

@article{fasano_multidimensional_1987,
 adsurl = {https://ui.adsabs.harvard.edu/abs/1987MNRAS.225..155F},
 author = {{Fasano}, G. and {Franceschini}, A.},
 doi = {10.1093/mnras/225.1.155},
 journal = {\mnras},
 month = {March},
 pages = {155-170},
 title = {{A multidimensional version of the Kolmogorov-Smirnov test}},
 volume = {225},
 year = {1987}
}

@article{fawcett_fundamental_2022,
 adsurl = {https://ui.adsabs.harvard.edu/abs/2022MNRAS.513.1254F},
 archiveprefix = {arXiv},
 author = {{Fawcett}, V.~A. and {Alexander}, D.~M. and {Rosario}, D.~J. and {Klindt}, L. and {Lusso}, E. and {Morabito}, L.~K. and {Calistro Rivera}, G.},
 doi = {10.1093/mnras/stac945},
 eprint = {2201.04139},
 journal = {\mnras},
 month = {June},
 number = {1},
 pages = {1254-1274},
 primaryclass = {astro-ph.GA},
 title = {{Fundamental differences in the properties of red and blue quasars: measuring the reddening and accretion properties with X-shooter}},
 volume = {513},
 year = {2022}
}

@article{filiz_ak_broad_2012,
 adsurl = {https://ui.adsabs.harvard.edu/abs/2012ApJ...757..114F},
 archiveprefix = {arXiv},
 author = {{Filiz Ak}, N. and {Brandt}, W.~N. and {Hall}, P.~B. and {Schneider}, D.~P. and {Anderson}, S.~F. and {Gibson}, R.~R. and {Lundgren}, B.~F. and {Myers}, A.~D. and {Petitjean}, P. and {Ross}, Nicholas P. and {Shen}, Yue and {York}, D.~G. and {Bizyaev}, D. and {Brinkmann}, J. and {Malanushenko}, E. and {Oravetz}, D.~J. and {Pan}, K. and {Simmons}, A.~E. and {Weaver}, B.~A.},
 doi = {10.1088/0004-637X/757/2/114},
 eid = {114},
 eprint = {1208.0836},
 journal = {\apj},
 month = {October},
 number = {2},
 pages = {114},
 primaryclass = {astro-ph.CO},
 title = {{Broad Absorption Line Disappearance on Multi-year Timescales in a Large Quasar Sample}},
 volume = {757},
 year = {2012}
}

@article{hamann_2018,
 adsurl = {https://ui.adsabs.harvard.edu/abs/2018MNRAS.476..943H},
 archiveprefix = {arXiv},
 author = {{Hamann}, Fred and {Chartas}, George and {Reeves}, James and {Nardini}, Emanuele},
 doi = {10.1093/mnras/sty043},
 eprint = {1801.04302},
 journal = {\mnras},
 month = {May},
 number = {1},
 pages = {943-953},
 primaryclass = {astro-ph.GA},
 title = {{Does the X-ray outflow quasar PDS 456 have a UV outflow at 0.3c?}},
 volume = {476},
 year = {2018}
}

@article{Hardcastle23_lofar_two-metre,
 adsurl = {https://ui.adsabs.harvard.edu/abs/2023A&A...678A.151H},
 archiveprefix = {arXiv},
 author = {{Hardcastle}, M.~J. and {Horton}, M.~A. and {Williams}, W.~L. and {Duncan}, K.~J. and {Alegre}, L. and {Barkus}, B. and {Croston}, J.~H. and {Dickinson}, H. and {Osinga}, E. and {R{\"o}ttgering}, H.~J.~A. and {Sabater}, J. and {Shimwell}, T.~W. and {Smith}, D.~J.~B. and {Best}, P.~N. and {Botteon}, A. and {Br{\"u}ggen}, M. and {Drabent}, A. and {de Gasperin}, F. and {G{\"u}rkan}, G. and {Hajduk}, M. and {Hale}, C.~L. and {Hoeft}, M. and {Jamrozy}, M. and {Kunert-Bajraszewska}, M. and {Kondapally}, R. and {Magliocchetti}, M. and {Mahatma}, V.~H. and {Mostert}, R.~I.~J. and {O'Sullivan}, S.~P. and {Pajdosz-{\'S}mierciak}, U. and {Petley}, J. and {Pierce}, J.~C.~S. and {Prandoni}, I. and {Schwarz}, D.~J. and {Shulewski}, A. and {Siewert}, T.~M. and {Stott}, J.~P. and {Tang}, H. and {Vaccari}, M. and {Zheng}, X. and {Bailey}, T. and {Desbled}, S. and {Goyal}, A. and {Gonano}, V. and {Hanset}, M. and {Kurtz}, W. and {Lim}, S.~M. and {Mielle}, L. and {Molloy}, C.~S. and {Roth}, R. and {Terentev}, I.~A. and {Torres}, M.},
 doi = {10.1051/0004-6361/202347333},
 eid = {A151},
 eprint = {2309.00102},
 journal = {\aap},
 month = {October},
 pages = {A151},
 primaryclass = {astro-ph.GA},
 title = {{The LOFAR Two-Metre Sky Survey. VI. Optical identifications for the second data release}},
 volume = {678},
 year = {2023}
}

@article{hidalgo_connection_2022,
 adsurl = {https://ui.adsabs.harvard.edu/abs/2022ApJ...939L..24R},
 archiveprefix = {arXiv},
 author = {{Rodr{\'\i}guez Hidalgo}, Paola and {Rankine}, Amy L.},
 doi = {10.3847/2041-8213/ac9628},
 eid = {L24},
 eprint = {2209.13642},
 journal = {\apjl},
 month = {November},
 number = {2},
 pages = {L24},
 primaryclass = {astro-ph.GA},
 title = {{Connection between Emission and Absorption Outflows through the Study of Quasars with Extremely High Velocity Outflows}},
 volume = {939},
 year = {2022}
}

@article{hidalgo_survey_2020,
 adsurl = {https://ui.adsabs.harvard.edu/abs/2020ApJ...896..151R},
 archiveprefix = {arXiv},
 author = {{Rodr{\'\i}guez Hidalgo}, Paola and {Khatri}, Abdul Moiz and {Hall}, Patrick B. and {Haas}, Sean and {Quintero}, Carla and {Khatu}, Viraja and {Kowash}, Griffin and {Murray}, Norm},
 doi = {10.3847/1538-4357/ab9198},
 eid = {151},
 eprint = {2006.05633},
 journal = {\apj},
 month = {June},
 number = {2},
 pages = {151},
 primaryclass = {astro-ph.GA},
 title = {{Survey of Extremely High-velocity Outflows in Sloan Digital Sky Survey Quasars}},
 volume = {896},
 year = {2020}
}

@article{hopkins_multi-phase_2024,
 adsurl = {https://ui.adsabs.harvard.edu/abs/2025OJAp....8E..56H},
 archiveprefix = {arXiv},
 author = {{Hopkins}, Philip F.},
 doi = {10.33232/001c.137969},
 eid = {56},
 eprint = {2407.00160},
 journal = {The Open Journal of Astrophysics},
 month = {May},
 pages = {56},
 primaryclass = {astro-ph.GA},
 title = {{Multi-Phase Thermal Structure \& The Origin of the Broad-Line Region, Torus, and Corona in Magnetically-Dominated Accretion Disks}},
 volume = {8},
 year = {2025}
}

@article{jackson_exploring_2026,
 adsurl = {https://ui.adsabs.harvard.edu/abs/2026MNRAS.546ag065J},
 archiveprefix = {arXiv},
 author = {{Jackson}, Charlotte L. and {Matthews}, James H. and {Whittam}, Imogen H. and {Jarvis}, Matt J. and {Temple}, Matthew J. and {Rankine}, Amy L. and {Hewett}, Paul C.},
 doi = {10.1093/mnras/stag065},
 eid = {stag065},
 eprint = {2510.25833},
 journal = {\mnras},
 month = {March},
 number = {3},
 pages = {stag065},
 primaryclass = {astro-ph.GA},
 title = {{Exploring the quasar disc─wind─jet connection with LoTSS and SDSS}},
 volume = {546},
 year = {2026}
}

@article{Kellermann89_vla_observations,
 adsurl = {https://ui.adsabs.harvard.edu/abs/1989AJ.....98.1195K},
 author = {{Kellermann}, K.~I. and {Sramek}, R. and {Schmidt}, M. and {Shaffer}, D.~B. and {Green}, R.},
 doi = {10.1086/115207},
 journal = {\aj},
 month = {October},
 pages = {1195},
 title = {{VLA Observations of Objects in the Palomar Bright Quasar Survey}},
 volume = {98},
 year = {1989}
}

@misc{kimball_2018_3942728,
 author = {Kimball, Amy E},
 doi = {10.5281/zenodo.3942728},
 month = {May},
 publisher = {Zenodo},
 title = {The origins of radio emission from [radio-"quiet"]
AGN
},
 url = {https://doi.org/10.5281/zenodo.3942728},
 year = {2018}
}

@article{Laor08_origin_radio,
 adsurl = {https://ui.adsabs.harvard.edu/abs/2008MNRAS.390..847L},
 archiveprefix = {arXiv},
 author = {{Laor}, Ari and {Behar}, Ehud},
 doi = {10.1111/j.1365-2966.2008.13806.x},
 eprint = {0808.0637},
 journal = {\mnras},
 month = {October},
 number = {2},
 pages = {847-862},
 primaryclass = {astro-ph},
 title = {{On the origin of radio emission in radio-quiet quasars}},
 volume = {390},
 year = {2008}
}

@article{Matthews16_testing_quasar,
 adsurl = {https://ui.adsabs.harvard.edu/abs/2016MNRAS.458..293M},
 archiveprefix = {arXiv},
 author = {{Matthews}, J.~H. and {Knigge}, C. and {Long}, K.~S. and {Sim}, S.~A. and {Higginbottom}, N. and {Mangham}, S.~W.},
 doi = {10.1093/mnras/stw323},
 eprint = {1602.02765},
 journal = {\mnras},
 month = {May},
 number = {1},
 pages = {293-305},
 primaryclass = {astro-ph.GA},
 title = {{Testing quasar unification: radiative transfer in clumpy winds}},
 volume = {458},
 year = {2016}
}

@article{Matthews23_disc_wind,
 adsurl = {https://ui.adsabs.harvard.edu/abs/2023MNRAS.526.3967M},
 archiveprefix = {arXiv},
 author = {{Matthews}, James H. and {Strong-Wright}, Jago and {Knigge}, Christian and {Hewett}, Paul and {Temple}, Matthew J. and {Long}, Knox S. and {Rankine}, Amy L. and {Stepney}, Matthew and {Banerji}, Manda and {Richards}, Gordon T.},
 doi = {10.1093/mnras/stad2895},
 eprint = {2309.14434},
 journal = {\mnras},
 month = {December},
 number = {3},
 pages = {3967-3986},
 primaryclass = {astro-ph.GA},
 title = {{A disc wind model for blueshifts in quasar broad emission lines}},
 volume = {526},
 year = {2023}
}

@article{Matthews26_how_massive,
 adsurl = {https://ui.adsabs.harvard.edu/abs/2026MNRAS.551g1424M},
 archiveprefix = {arXiv},
 author = {{Matthews}, James H.},
 doi = {10.1093/mnras/stag1424},
 eid = {stag1424},
 eprint = {2607.25035},
 journal = {\mnras},
 month = {September},
 number = {1},
 pages = {stag1424},
 primaryclass = {astro-ph.GA},
 title = {{How massive and clumpy must a quasar wind be to create emission line blueshifts?}},
 volume = {551},
 year = {2026}
}

@article{matzeu_supermassive_2023,
 adsurl = {https://ui.adsabs.harvard.edu/abs/2023A&A...670A.182M},
 archiveprefix = {arXiv},
 author = {{Matzeu}, G.~A. and {Brusa}, M. and {Lanzuisi}, G. and {Dadina}, M. and {Bianchi}, S. and {Kriss}, G. and {Mehdipour}, M. and {Nardini}, E. and {Chartas}, G. and {Middei}, R. and {Piconcelli}, E. and {Gianolli}, V. and {Comastri}, A. and {Longinotti}, A.~L. and {Krongold}, Y. and {Ricci}, F. and {Petrucci}, P.~O. and {Tombesi}, F. and {Luminari}, A. and {Zappacosta}, L. and {Miniutti}, G. and {Gaspari}, M. and {Behar}, E. and {Bischetti}, M. and {Mathur}, S. and {Perna}, M. and {Giustini}, M. and {Grandi}, P. and {Torresi}, E. and {Vignali}, C. and {Bruni}, G. and {Cappi}, M. and {Costantini}, E. and {Cresci}, G. and {De Marco}, B. and {De Rosa}, A. and {Gilli}, R. and {Guainazzi}, M. and {Kaastra}, J. and {Kraemer}, S. and {La Franca}, F. and {Marconi}, A. and {Panessa}, F. and {Ponti}, G. and {Proga}, D. and {Ursini}, F. and {Baldini}, P. and {Fiore}, F. and {King}, A.~R. and {Maiolino}, R. and {Matt}, G. and {Merloni}, A.},
 doi = {10.1051/0004-6361/202245036},
 eid = {A182},
 eprint = {2212.02960},
 journal = {\aap},
 month = {February},
 pages = {A182},
 primaryclass = {astro-ph.HE},
 title = {{Supermassive Black Hole Winds in X-rays: SUBWAYS. I. Ultra-fast outflows in quasars beyond the local Universe}},
 volume = {670},
 year = {2023}
}

@article{Morabito25_decade_sub-arcsecond,
 adsurl = {https://ui.adsabs.harvard.edu/abs/2025Ap&SS.370...19M},
 archiveprefix = {arXiv},
 author = {{Morabito}, Leah K. and {Jackson}, Neal and {de Jong}, Jurjen and {Escott}, Emmy and {Groeneveld}, Christian and {Mahatma}, Vijay and {Petley}, James and {Sweijen}, Frits and {Timmerman}, Roland and {van Weeren}, Reinout J.},
 doi = {10.1007/s10509-025-04406-x},
 eid = {19},
 eprint = {2502.06946},
 journal = {\apss},
 month = {February},
 number = {2},
 pages = {19},
 primaryclass = {astro-ph.IM},
 title = {{A decade of sub-arcsecond imaging with the International LOFAR Telescope}},
 volume = {370},
 year = {2025}
}

@article{morabito_origin_2019,
 adsurl = {https://ui.adsabs.harvard.edu/abs/2019A&A...622A..15M},
 archiveprefix = {arXiv},
 author = {{Morabito}, L.~K. and {Matthews}, J.~H. and {Best}, P.~N. and {G{\"u}rkan}, G. and {Jarvis}, M.~J. and {Prandoni}, I. and {Duncan}, K.~J. and {Hardcastle}, M.~J. and {Kunert-Bajraszewska}, M. and {Mechev}, A.~P. and {Mooney}, S. and {Sabater}, J. and {R{\"o}ttgering}, H.~J.~A. and {Shimwell}, T.~W. and {Smith}, D.~J.~B. and {Tasse}, C. and {Williams}, W.~L.},
 doi = {10.1051/0004-6361/201833821},
 eid = {A15},
 eprint = {1811.07931},
 journal = {\aap},
 month = {February},
 pages = {A15},
 primaryclass = {astro-ph.GA},
 title = {{The origin of radio emission in broad absorption line quasars: Results from the LOFAR Two-metre Sky Survey}},
 volume = {622},
 year = {2019}
}

@article{Nims15_observational_signatures,
 adsurl = {https://ui.adsabs.harvard.edu/abs/2015MNRAS.447.3612N},
 archiveprefix = {arXiv},
 author = {{Nims}, Jesse and {Quataert}, Eliot and {Faucher-Gigu{\`e}re}, Claude-Andr{\'e}},
 doi = {10.1093/mnras/stu2648},
 eprint = {1408.5141},
 journal = {\mnras},
 month = {March},
 number = {4},
 pages = {3612-3622},
 primaryclass = {astro-ph.GA},
 title = {{Observational signatures of galactic winds powered by active galactic nuclei}},
 volume = {447},
 year = {2015}
}

@article{Panessa19_origin_radio,
 adsurl = {https://ui.adsabs.harvard.edu/abs/2019NatAs...3..387P},
 archiveprefix = {arXiv},
 author = {{Panessa}, Francesca and {Baldi}, Ranieri Diego and {Laor}, Ari and {Padovani}, Paolo and {Behar}, Ehud and {McHardy}, Ian},
 doi = {10.1038/s41550-019-0765-4},
 eprint = {1902.05917},
 journal = {Nature Astronomy},
 month = {April},
 pages = {387-396},
 primaryclass = {astro-ph.GA},
 title = {{The origin of radio emission from radio-quiet active galactic nuclei}},
 volume = {3},
 year = {2019}
}

@article{Paris12_sloan_digital,
 adsurl = {https://ui.adsabs.harvard.edu/abs/2012A&A...548A..66P},
 archiveprefix = {arXiv},
 author = {{P{\^a}ris}, I. and {Petitjean}, P. and {Aubourg}, {\'E}. and {Bailey}, S. and {Ross}, N.~P. and {Myers}, A.~D. and {Strauss}, M.~A. and {Anderson}, S.~F. and {Arnau}, E. and {Bautista}, J. and {Bizyaev}, D. and {Bolton}, A.~S. and {Bovy}, J. and {Brandt}, W.~N. and {Brewington}, H. and {Browstein}, J.~R. and {Busca}, N. and {Capellupo}, D. and {Carithers}, W. and {Croft}, R.~A.~C. and {Dawson}, K. and {Delubac}, T. and {Ebelke}, G. and {Eisenstein}, D.~J. and {Engelke}, P. and {Fan}, X. and {Filiz Ak}, N. and {Finley}, H. and {Font-Ribera}, A. and {Ge}, J. and {Gibson}, R.~R. and {Hall}, P.~B. and {Hamann}, F. and {Hennawi}, J.~F. and {Ho}, S. and {Hogg}, D.~W. and {Ivezi{\'c}}, {\v{Z}}. and {Jiang}, L. and {Kimball}, A.~E. and {Kirkby}, D. and {Kirkpatrick}, J.~A. and {Lee}, K.-G. and {Le Goff}, J.-M. and {Lundgren}, B. and {MacLeod}, C.~L. and {Malanushenko}, E. and {Malanushenko}, V. and {Maraston}, C. and {McGreer}, I.~D. and {McMahon}, R.~G. and {Miralda-Escud{\'e}}, J. and {Muna}, D. and {Noterdaeme}, P. and {Oravetz}, D. and {Palanque-Delabrouille}, N. and {Pan}, K. and {Perez-Fournon}, I. and {Pieri}, M.~M. and {Richards}, G.~T. and {Rollinde}, E. and {Sheldon}, E.~S. and {Schlegel}, D.~J. and {Schneider}, D.~P. and {Slosar}, A. and {Shelden}, A. and {Shen}, Y. and {Simmons}, A. and {Snedden}, S. and {Suzuki}, N. and {Tinker}, J. and {Viel}, M. and {Weaver}, B.~A. and {Weinberg}, D.~H. and {White}, M. and {Wood-Vasey}, W.~M. and {Y{\`e}che}, C.},
 doi = {10.1051/0004-6361/201220142},
 eid = {A66},
 eprint = {1210.5166},
 journal = {\aap},
 month = {December},
 pages = {A66},
 primaryclass = {astro-ph.CO},
 title = {{The Sloan Digital Sky Survey quasar catalog: ninth data release}},
 volume = {548},
 year = {2012}
}

@article{peacock_two-dimensional_1983,
 adsurl = {https://ui.adsabs.harvard.edu/abs/1983MNRAS.202..615P},
 author = {{Peacock}, J.~A.},
 doi = {10.1093/mnras/202.3.615},
 journal = {\mnras},
 month = {February},
 pages = {615-627},
 title = {{Two-dimensional goodness-of-fit testing in astronomy.}},
 volume = {202},
 year = {1983}
}

@article{petley_connecting_2022,
 adsurl = {https://ui.adsabs.harvard.edu/abs/2022MNRAS.515.5159P},
 archiveprefix = {arXiv},
 author = {{Petley}, J.~W. and {Morabito}, L.~K. and {Alexander}, D.~M. and {Rankine}, A.~L. and {Fawcett}, V.~A. and {Rosario}, D.~J. and {Matthews}, J.~H. and {Shimwell}, T.~M. and {Drabent}, A.},
 doi = {10.1093/mnras/stac2067},
 eprint = {2207.10102},
 journal = {\mnras},
 month = {October},
 number = {4},
 pages = {5159-5174},
 primaryclass = {astro-ph.GA},
 title = {{Connecting radio emission to AGN wind properties with broad absorption line quasars}},
 volume = {515},
 year = {2022}
}

@article{petley_how_2024,
 adsurl = {https://ui.adsabs.harvard.edu/abs/2024MNRAS.529.1995P},
 archiveprefix = {arXiv},
 author = {{Petley}, James W. and {Morabito}, Leah K. and {Rankine}, Amy L. and {Richards}, Gordon T. and {Thomas}, Nicole L. and {Alexander}, David M. and {Fawcett}, Victoria A. and {Calistro Rivera}, Gabriela and {Prandoni}, Isabella and {Best}, Philip N. and {Kolwa}, Sthabile},
 doi = {10.1093/mnras/stae626},
 eprint = {2402.18623},
 journal = {\mnras},
 month = {April},
 number = {3},
 pages = {1995-2007},
 primaryclass = {astro-ph.GA},
 title = {{How does the radio enhancement of broad absorption line quasars relate to colour and accretion rate?}},
 volume = {529},
 year = {2024}
}

@article{PRH_2011,
 adsurl = {https://ui.adsabs.harvard.edu/abs/2011MNRAS.411..247R},
 archiveprefix = {arXiv},
 author = {{Rodr{\'\i}guez Hidalgo}, Paola and {Hamann}, Fred and {Hall}, Patrick},
 doi = {10.1111/j.1365-2966.2010.17677.x},
 eprint = {1009.1890},
 journal = {\mnras},
 month = {February},
 number = {1},
 pages = {247-259},
 primaryclass = {astro-ph.CO},
 title = {{The extremely high velocity outflow in quasar PG0935+417}},
 volume = {411},
 year = {2011}
}

@article{PRH_2025,
 adsurl = {https://ui.adsabs.harvard.edu/abs/2025ApJ...990..152R},
 archiveprefix = {arXiv},
 author = {{Rodr{\'\i}guez Hidalgo}, Paola and {Choi}, Hyunseop and {Hall}, Patrick B. and {Leighly}, Karen M. and {Flores}, Liliana and {Charles}, Mikel M. and {DeFrancesco}, Cora and {Hlavacek-Larrondo}, Julie and {Perreault-Levasseur}, Laurence},
 doi = {10.3847/1538-4357/adef0d},
 eid = {152},
 eprint = {2508.14221},
 journal = {\apj},
 month = {September},
 number = {2},
 pages = {152},
 primaryclass = {astro-ph.GA},
 title = {{Massive Extremely High-velocity Outflow in the Quasar J164653.72+243942.2}},
 volume = {990},
 year = {2025}
}

@article{Rankine20_bal_non-bal,
 adsurl = {https://ui.adsabs.harvard.edu/abs/2020MNRAS.492.4553R},
 archiveprefix = {arXiv},
 author = {{Rankine}, Amy L. and {Hewett}, Paul C. and {Banerji}, Manda and {Richards}, Gordon T.},
 doi = {10.1093/mnras/staa130},
 eprint = {1912.08700},
 journal = {\mnras},
 month = {March},
 number = {3},
 pages = {4553-4575},
 primaryclass = {astro-ph.GA},
 title = {{BAL and non-BAL quasars: continuum, emission, and absorption properties establish a common parent sample}},
 volume = {492},
 year = {2020}
}

@article{rankine_placing_2021,
 adsurl = {https://ui.adsabs.harvard.edu/abs/2021MNRAS.502.4154R},
 archiveprefix = {arXiv},
 author = {{Rankine}, Amy L. and {Matthews}, James H. and {Hewett}, Paul C. and {Banerji}, Manda and {Morabito}, Leah K. and {Richards}, Gordon T.},
 doi = {10.1093/mnras/stab302},
 eprint = {2101.12635},
 journal = {\mnras},
 month = {April},
 number = {3},
 pages = {4154-4169},
 primaryclass = {astro-ph.GA},
 title = {{Placing LOFAR-detected quasars in C IV emission space: implications for winds, jets and star formation}},
 volume = {502},
 year = {2021}
}

@article{Richards21_probing_wind,
 adsurl = {https://ui.adsabs.harvard.edu/abs/2021AJ....162..270R},
 archiveprefix = {arXiv},
 author = {{Richards}, Gordon T. and {McCaffrey}, Trevor V. and {Kimball}, Amy and {Rankine}, Amy L. and {Matthews}, James H. and {Hewett}, Paul C. and {Rivera}, Angelica B.},
 doi = {10.3847/1538-3881/ac283b},
 eid = {270},
 eprint = {2106.07783},
 journal = {\aj},
 month = {December},
 number = {6},
 pages = {270},
 primaryclass = {astro-ph.GA},
 title = {{Probing the Wind Component of Radio Emission in Luminous High-redshift Quasars}},
 volume = {162},
 year = {2021}
}

@article{richards_spectral_2006,
 adsurl = {https://ui.adsabs.harvard.edu/abs/2006ApJS..166..470R},
 archiveprefix = {arXiv},
 author = {{Richards}, Gordon T. and {Lacy}, Mark and {Storrie-Lombardi}, Lisa J. and {Hall}, Patrick B. and {Gallagher}, S.~C. and {Hines}, Dean C. and {Fan}, Xiaohui and {Papovich}, Casey and {Vanden Berk}, Daniel E. and {Trammell}, George B. and {Schneider}, Donald P. and {Vestergaard}, Marianne and {York}, Donald G. and {Jester}, Sebastian and {Anderson}, Scott F. and {Budav{\'a}ri}, Tam{\'a}s and {Szalay}, Alexander S.},
 doi = {10.1086/506525},
 eprint = {astro-ph/0601558},
 journal = {\apjs},
 month = {October},
 number = {2},
 pages = {470-497},
 primaryclass = {astro-ph},
 title = {{Spectral Energy Distributions and Multiwavelength Selection of Type 1 Quasars}},
 volume = {166},
 year = {2006}
}

@article{scikit-learn,
 author = {Pedregosa, F. and Varoquaux, G. and Gramfort, A. and Michel, V.
and Thirion, B. and Grisel, O. and Blondel, M. and Prettenhofer, P.
and Weiss, R. and Dubourg, V. and Vanderplas, J. and Passos, A. and
Cournapeau, D. and Brucher, M. and Perrot, M. and Duchesnay, E.},
 journal = {Journal of Machine Learning Research},
 pages = {2825--2830},
 title = {Scikit-learn: Machine Learning in {P}ython},
 volume = {12},
 year = {2011}
}

@article{seaton_new_2026,
 adsurl = {https://ui.adsabs.harvard.edu/abs/2026ApJ..1004...49S},
 archiveprefix = {arXiv},
 author = {{Seaton}, Lucas M. and {Hall}, Patrick B. and {Flores}, Liliana and {Rodr{\'\i}guez Hidalgo}, Paola and {Veltri}, Marianna and {Zhu}, Zezhou and {Serna}, Javier and {Brandt}, W. Niel and {Anderson}, Scott and {Assef}, Roberto J. and {Ba{\~n}ados}, Eduardo and {Grier}, Catherine J. and {Homayouni}, Yasaman and {Morrison}, Sean and {Negrete}, C. Alenka and {Rankine}, Amy L. and {Runnoe}, Jessie and {Schneider}, Donald P. and {Shen}, Yue and {Temple}, Matthew and {Trakhtenbrot}, Benny and {Trump}, Jonathan R. and {Weiss}, Erik},
 doi = {10.3847/1538-4357/ae5f94},
 eid = {49},
 eprint = {2606.06226},
 journal = {\apj},
 month = {June},
 number = {1},
 pages = {49},
 primaryclass = {astro-ph.GA},
 title = {{A New Member of the Fast and Furious Family: A Relativistic and Time-variable UV Outflow in a Luminous Quasar}},
 volume = {1004},
 year = {2026}
}

@article{shimwell_lofar_2026,
 adsurl = {https://ui.adsabs.harvard.edu/abs/2026A&A...707A.198S},
 archiveprefix = {arXiv},
 author = {{Shimwell}, T.~W. and {Hardcastle}, M.~J. and {Tasse}, C. and {Drabent}, A. and {Botteon}, A. and {Williams}, W.~L. and {Best}, P.~N. and {R{\"o}ttgering}, H.~J.~A. and {Br{\"u}ggen}, M. and {Brunetti}, G. and {Callingham}, J.~R. and {Chy{\.z}y}, K.~T. and {Conway}, J.~E. and {De Gasperin}, F. and {Haverkorn}, M. and {Horellou}, C. and {Jackson}, N. and {Miley}, G.~K. and {Morabito}, L.~K. and {Morganti}, R. and {O'Sullivan}, S.~P. and {Schwarz}, D.~J. and {Smith}, D.~J.~B. and {van Weeren}, R.~J. and {Vedantham}, H.~K. and {White}, G.~J. and {Ahmadi}, A. and {Alegre}, L. and {Arias}, M. and {Asabere}, B. and {Bahr-Kalus}, B. and {Barkus}, B. and {Bilicki}, M. and {B{\"o}hme}, L. and {Brentjens}, M. and {Brienza}, M. and {Bomans}, D.~J. and {Bonafede}, A. and {Bonato}, M. and {Bonnassieux}, E. and {Boxelaar}, J.~M. and {Camera}, S. and {Cassano}, R. and {Chilufya}, J. and {Cianfaglione}, M. and {Croston}, J.~H. and {Cuciti}, V. and {Dabhade}, P. and {De Rubeis}, E. and {de Jong}, J.~M.~G.~H.~J. and {Dallacasa}, D. and {Dettmar}, R.~J. and {Duncan}, K.~J. and {Di Gennaro}, G. and {Edler}, H.~W. and {Groeneveld}, C. and {G{\"u}rkan}, G. and {Hajduk}, M. and {Hale}, C.~L. and {Heesen}, V. and {Hoang}, D.~N. and {Hoeft}, M. and {Holties}, H. and {Horton}, M.~A. and {Iacobelli}, M. and {Jamrozy}, M. and {Jarvis}, M.~J. and {Jelic}, V. and {Kadler}, M. and {Kondapally}, R. and {Kunert-Bajraszewska}, M. and {Loose}, M. and {Magliocchetti}, M. and {Ma{\l}ek}, K. and {Manzano}, C. and {McKean}, J.~P. and {Mevius}, M. and {Mingo}, B. and {Miskolczi}, A. and {Misra}, A. and {Mold{\'o}n}, J. and {Nair}, D.~G. and {Nakoneczny}, S.~J. and {Orru}, E. and {Pashapour-Ahmadabadi}, M. and {Pasini}, T. and {Petley}, J. and {Pierce}, J.~C.~S. and {Prandoni}, I. and {Rafferty}, D. and {Rajpurohit}, K. and {Riseley}, C.~J. and {Roberts}, I.~D. and {Sethi}, S. and {Shulevski}, A. and {Stein}, M. and {Stuardi}, C. and {Sweijen}, F. and {ter Veen}, S. and {Timmerman}, R. and {Vaccari}, M. and {Wijnholds}, S.},
 doi = {10.1051/0004-6361/202557749},
 eid = {A198},
 eprint = {2602.15949},
 journal = {\aap},
 month = {March},
 pages = {A198},
 primaryclass = {astro-ph.GA},
 title = {{The LOFAR Two-metre Sky Survey: VII. Third Data Release}},
 volume = {707},
 year = {2026}
}

@article{sotomayor_nonthermal_2022,
 adsurl = {https://ui.adsabs.harvard.edu/abs/2022A&A...664A.178S},
 archiveprefix = {arXiv},
 author = {{Sotomayor}, Pablo and {Romero}, Gustavo E.},
 doi = {10.1051/0004-6361/202243682},
 eid = {A178},
 eprint = {2206.10731},
 journal = {\aap},
 month = {August},
 pages = {A178},
 primaryclass = {astro-ph.HE},
 title = {{Nonthermal radiation from the central region of super-accreting active galactic nuclei}},
 volume = {664},
 year = {2022}
}

@article{Strittmatter80_radio_observations,
 adsurl = {https://ui.adsabs.harvard.edu/abs/1980A&A....88L..12S},
 author = {{Strittmatter}, P.~A. and {Hill}, P. and {Pauliny-Toth}, I.~I.~K. and {Steppe}, H. and {Witzel}, A.},
 journal = {\aap},
 month = {August},
 number = {3},
 pages = {L12-L15},
 title = {{Radio observations of optically selected quasars}},
 volume = {88},
 year = {1980}
}

@article{Temple23_testing_agn,
 adsurl = {https://ui.adsabs.harvard.edu/abs/2023MNRAS.523..646T},
 archiveprefix = {arXiv},
 author = {{Temple}, Matthew J. and {Matthews}, James H. and {Hewett}, Paul C. and {Rankine}, Amy L. and {Richards}, Gordon T. and {Banerji}, Manda and {Ferland}, Gary J. and {Knigge}, Christian and {Stepney}, Matthew},
 doi = {10.1093/mnras/stad1448},
 eprint = {2301.02675},
 journal = {\mnras},
 month = {July},
 number = {1},
 pages = {646-666},
 primaryclass = {astro-ph.GA},
 title = {{Testing AGN outflow and accretion models with C IV and He II emission line demographics in z {\ensuremath{\approx}} 2 quasars}},
 volume = {523},
 year = {2023}
}

@article{tombesi_evidence_2010,
 adsurl = {https://ui.adsabs.harvard.edu/abs/2010A&A...521A..57T},
 archiveprefix = {arXiv},
 author = {{Tombesi}, F. and {Cappi}, M. and {Reeves}, J.~N. and {Palumbo}, G.~G.~C. and {Yaqoob}, T. and {Braito}, V. and {Dadina}, M.},
 doi = {10.1051/0004-6361/200913440},
 eid = {A57},
 eprint = {1006.2858},
 journal = {\aap},
 month = {October},
 pages = {A57},
 primaryclass = {astro-ph.HE},
 title = {{Evidence for ultra-fast outflows in radio-quiet AGNs. I. Detection and statistical incidence of Fe K-shell absorption lines}},
 volume = {521},
 year = {2010}
}

@article{van_haarlem_lofar_2013,
 adsurl = {https://ui.adsabs.harvard.edu/abs/2013A&A...556A...2V},
 archiveprefix = {arXiv},
 author = {{van Haarlem}, M.~P. and {Wise}, M.~W. and {Gunst}, A.~W. and {Heald}, G. and {McKean}, J.~P. and {Hessels}, J.~W.~T. and {de Bruyn}, A.~G. and {Nijboer}, R. and {Swinbank}, J. and {Fallows}, R. and {Brentjens}, M. and {Nelles}, A. and {Beck}, R. and {Falcke}, H. and {Fender}, R. and {H{\"o}randel}, J. and {Koopmans}, L.~V.~E. and {Mann}, G. and {Miley}, G. and {R{\"o}ttgering}, H. and {Stappers}, B.~W. and {Wijers}, R.~A.~M.~J. and {Zaroubi}, S. and {van den Akker}, M. and {Alexov}, A. and {Anderson}, J. and {Anderson}, K. and {van Ardenne}, A. and {Arts}, M. and {Asgekar}, A. and {Avruch}, I.~M. and {Batejat}, F. and {B{\"a}hren}, L. and {Bell}, M.~E. and {Bell}, M.~R. and {van Bemmel}, I. and {Bennema}, P. and {Bentum}, M.~J. and {Bernardi}, G. and {Best}, P. and {B{\^\i}rzan}, L. and {Bonafede}, A. and {Boonstra}, A.-J. and {Braun}, R. and {Bregman}, J. and {Breitling}, F. and {van de Brink}, R.~H. and {Broderick}, J. and {Broekema}, P.~C. and {Brouw}, W.~N. and {Br{\"u}ggen}, M. and {Butcher}, H.~R. and {van Cappellen}, W. and {Ciardi}, B. and {Coenen}, T. and {Conway}, J. and {Coolen}, A. and {Corstanje}, A. and {Damstra}, S. and {Davies}, O. and {Deller}, A.~T. and {Dettmar}, R.-J. and {van Diepen}, G. and {Dijkstra}, K. and {Donker}, P. and {Doorduin}, A. and {Dromer}, J. and {Drost}, M. and {van Duin}, A. and {Eisl{\"o}ffel}, J. and {van Enst}, J. and {Ferrari}, C. and {Frieswijk}, W. and {Gankema}, H. and {Garrett}, M.~A. and {de Gasperin}, F. and {Gerbers}, M. and {de Geus}, E. and {Grie{\ss}meier}, J.-M. and {Grit}, T. and {Gruppen}, P. and {Hamaker}, J.~P. and {Hassall}, T. and {Hoeft}, M. and {Holties}, H.~A. and {Horneffer}, A. and {van der Horst}, A. and {van Houwelingen}, A. and {Huijgen}, A. and {Iacobelli}, M. and {Intema}, H. and {Jackson}, N. and {Jelic}, V. and {de Jong}, A. and {Juette}, E. and {Kant}, D. and {Karastergiou}, A. and {Koers}, A. and {Kollen}, H. and {Kondratiev}, V.~I. and {Kooistra}, E. and {Koopman}, Y. and {Koster}, A. and {Kuniyoshi}, M. and {Kramer}, M. and {Kuper}, G. and {Lambropoulos}, P. and {Law}, C. and {van Leeuwen}, J. and {Lemaitre}, J. and {Loose}, M. and {Maat}, P. and {Macario}, G. and {Markoff}, S. and {Masters}, J. and {McFadden}, R.~A. and {McKay-Bukowski}, D. and {Meijering}, H. and {Meulman}, H. and {Mevius}, M. and {Middelberg}, E. and {Millenaar}, R. and {Miller-Jones}, J.~C.~A. and {Mohan}, R.~N. and {Mol}, J.~D. and {Morawietz}, J. and {Morganti}, R. and {Mulcahy}, D.~D. and {Mulder}, E. and {Munk}, H. and {Nieuwenhuis}, L. and {van Nieuwpoort}, R. and {Noordam}, J.~E. and {Norden}, M. and {Noutsos}, A. and {Offringa}, A.~R. and {Olofsson}, H. and {Omar}, A. and {Orr{\'u}}, E. and {Overeem}, R. and {Paas}, H. and {Pandey-Pommier}, M. and {Pandey}, V.~N. and {Pizzo}, R. and {Polatidis}, A. and {Rafferty}, D. and {Rawlings}, S. and {Reich}, W. and {de Reijer}, J.-P. and {Reitsma}, J. and {Renting}, G.~A. and {Riemers}, P. and {Rol}, E. and {Romein}, J.~W. and {Roosjen}, J. and {Ruiter}, M. and {Scaife}, A. and {van der Schaaf}, K. and {Scheers}, B. and {Schellart}, P. and {Schoenmakers}, A. and {Schoonderbeek}, G. and {Serylak}, M. and {Shulevski}, A. and {Sluman}, J. and {Smirnov}, O. and {Sobey}, C. and {Spreeuw}, H. and {Steinmetz}, M. and {Sterks}, C.~G.~M. and {Stiepel}, H.-J. and {Stuurwold}, K. and {Tagger}, M. and {Tang}, Y. and {Tasse}, C. and {Thomas}, I. and {Thoudam}, S. and {Toribio}, M.~C. and {van der Tol}, B. and {Usov}, O. and {van Veelen}, M. and {van der Veen}, A.-J. and {ter Veen}, S. and {Verbiest}, J.~P.~W. and {Vermeulen}, R. and {Vermaas}, N. and {Vocks}, C. and {Vogt}, C. and {de Vos}, M. and {van der Wal}, E. and {van Weeren}, R. and {Weggemans}, H. and {Weltevrede}, P. and {White}, S. and {Wijnholds}, S.~J. and {Wilhelmsson}, T. and {Wucknitz}, O. and {Yatawatta}, S. and {Zarka}, P. and {Zensus}, A.},
 doi = {10.1051/0004-6361/201220873},
 eid = {A2},
 eprint = {1305.3550},
 journal = {\aap},
 month = {August},
 pages = {A2},
 primaryclass = {astro-ph.IM},
 title = {{LOFAR: The LOw-Frequency ARray}},
 volume = {556},
 year = {2013}
}

@article{vestergaard_determining_2006,
 adsurl = {https://ui.adsabs.harvard.edu/abs/2006ApJ...641..689V},
 archiveprefix = {arXiv},
 author = {{Vestergaard}, Marianne and {Peterson}, Bradley M.},
 doi = {10.1086/500572},
 eprint = {astro-ph/0601303},
 journal = {\apj},
 month = {April},
 number = {2},
 pages = {689-709},
 primaryclass = {astro-ph},
 title = {{Determining Central Black Hole Masses in Distant Active Galaxies and Quasars. II. Improved Optical and UV Scaling Relationships}},
 volume = {641},
 year = {2006}
}

@article{Vivek26_recurrent_multiyear,
 adsurl = {https://ui.adsabs.harvard.edu/abs/2026ApJ..1007..113V},
 author = {{Vivek}, M. and {Aromal}, P. and {Srianand}, R. and {Anjali}, K.~A. and {Gallagher}, S.~C. and {Rachana}},
 doi = {10.3847/1538-4357/ae8aec},
 eid = {113},
 journal = {\apj},
 month = {August},
 number = {2},
 pages = {113},
 title = {{Recurrent Multiyear Mg II Broad Absorption Line Variability in SDSS J1333+0012}},
 volume = {1007},
 year = {2026}
}

@article{Wang18,
 adsurl = {https://ui.adsabs.harvard.edu/abs/2018ApJ...869L...9W},
 archiveprefix = {arXiv},
 author = {{Wang}, Feige and {Yang}, Jinyi and {Fan}, Xiaohui and {Yue}, Minghao and {Wu}, Xue-Bing and {Schindler}, Jan-Torge and {Bian}, Fuyan and {Li}, Jiang-Tao and {Farina}, Emanuele P. and {Ba{\~n}ados}, Eduardo and {Davies}, Frederick B. and {Decarli}, Roberto and {Green}, Richard and {Jiang}, Linhua and {Hennawi}, Joseph F. and {Huang}, Yun-Hsin and {Mazzucchelli}, Chiara and {McGreer}, Ian D. and {Venemans}, Bram and {Walter}, Fabian and {Beletsky}, Yuri},
 doi = {10.3847/2041-8213/aaf1d2},
 eid = {L9},
 eprint = {1810.11925},
 journal = {\apjl},
 month = {December},
 number = {1},
 pages = {L9},
 primaryclass = {astro-ph.GA},
 title = {{The Discovery of a Luminous Broad Absorption Line Quasar at a Redshift of 7.02}},
 volume = {869},
 year = {2018}
}

@article{Wang21,
 adsurl = {https://ui.adsabs.harvard.edu/abs/2021ApJ...907L...1W},
 archiveprefix = {arXiv},
 author = {{Wang}, Feige and {Yang}, Jinyi and {Fan}, Xiaohui and {Hennawi}, Joseph F. and {Barth}, Aaron J. and {Banados}, Eduardo and {Bian}, Fuyan and {Boutsia}, Konstantina and {Connor}, Thomas and {Davies}, Frederick B. and {Decarli}, Roberto and {Eilers}, Anna-Christina and {Farina}, Emanuele Paolo and {Green}, Richard and {Jiang}, Linhua and {Li}, Jiang-Tao and {Mazzucchelli}, Chiara and {Nanni}, Riccardo and {Schindler}, Jan-Torge and {Venemans}, Bram and {Walter}, Fabian and {Wu}, Xue-Bing and {Yue}, Minghao},
 doi = {10.3847/2041-8213/abd8c6},
 eid = {L1},
 eprint = {2101.03179},
 journal = {\apjl},
 month = {January},
 number = {1},
 pages = {L1},
 primaryclass = {astro-ph.GA},
 title = {{A Luminous Quasar at Redshift 7.642}},
 volume = {907},
 year = {2021}
}

@article{weymann_comparisons_1991,
 adsurl = {https://ui.adsabs.harvard.edu/abs/1991ApJ...373...23W},
 author = {{Weymann}, Ray J. and {Morris}, Simon L. and {Foltz}, Craig B. and {Hewett}, Paul C.},
 doi = {10.1086/170020},
 journal = {\apj},
 month = {May},
 pages = {23},
 title = {{Comparisons of the Emission-Line and Continuum Properties of Broad Absorption Line and Normal Quasi-stellar Objects}},
 volume = {373},
 year = {1991}
}

@article{Wilkes84_studies_broad,
 adsurl = {https://ui.adsabs.harvard.edu/abs/1984MNRAS.207...73W},
 author = {{Wilkes}, B.~J.},
 doi = {10.1093/mnras/207.1.73},
 journal = {\mnras},
 month = {March},
 pages = {73-98},
 title = {{Studies of broad emission line profiles in QSOs - I. Observed, high-resolution profiles.}},
 volume = {207},
 year = {1984}
}

@article{yue_novel_2024,
 adsurl = {https://ui.adsabs.harvard.edu/abs/2024MNRAS.529.3939Y},
 archiveprefix = {arXiv},
 author = {{Yue}, B.-H. and {Best}, P.~N. and {Duncan}, K.~J. and {Calistro-Rivera}, G. and {Morabito}, L.~K. and {Petley}, J.~W. and {Prandoni}, I. and {R{\"o}ttgering}, H.~J.~A. and {Smith}, D.~J.~B.},
 doi = {10.1093/mnras/stae725},
 eprint = {2403.07074},
 journal = {\mnras},
 month = {April},
 number = {4},
 pages = {3939-3957},
 primaryclass = {astro-ph.GA},
 title = {{A novel Bayesian approach for decomposing the radio emission of quasars: I. Modelling the radio excess in red quasars}},
 volume = {529},
 year = {2024}
}

@article{yue_novel_2025,
 adsurl = {https://ui.adsabs.harvard.edu/abs/2025MNRAS.537..858Y},
 archiveprefix = {arXiv},
 author = {{Yue}, B.-H. and {Duncan}, K.~J. and {Best}, P.~N. and {Arnaudova}, M.~I. and {Morabito}, L.~K. and {Petley}, J.~W. and {R{\"o}ttgering}, H.~J.~A. and {Shenoy}, S. and {Smith}, D.~J.~B.},
 doi = {10.1093/mnras/staf077},
 eprint = {2501.07629},
 journal = {\mnras},
 month = {February},
 number = {2},
 pages = {858-875},
 primaryclass = {astro-ph.GA},
 title = {{A novel Bayesian approach for decomposing the radio emission of quasars - II. Link between quasar radio emission and black hole mass}},
 volume = {537},
 year = {2025}
}

@article{Zakamska14_quasar_feedback,
 adsurl = {https://ui.adsabs.harvard.edu/abs/2014MNRAS.442..784Z},
 archiveprefix = {arXiv},
 author = {{Zakamska}, Nadia L. and {Greene}, Jenny E.},
 doi = {10.1093/mnras/stu842},
 eprint = {1402.6736},
 journal = {\mnras},
 month = {July},
 number = {1},
 pages = {784-804},
 primaryclass = {astro-ph.GA},
 title = {{Quasar feedback and the origin of radio emission in radio-quiet quasars}},
 volume = {442},
 year = {2014}
}

@ARTICLE{Raginski16_agn_coronal,
       author = {{Raginski}, I. and {Laor}, Ari},
        title = "{AGN coronal emission models - I. The predicted radio emission}",
      journal = {\mnras},
         year = 2016,
        month = jun,
       volume = {459},
       number = {2},
        pages = {2082-2096},
          doi = {10.1093/mnras/stw772},
archivePrefix = {arXiv},
       eprint = {1604.00646},
 primaryClass = {astro-ph.GA},
       adsurl = {https://ui.adsabs.harvard.edu/abs/2016MNRAS.459.2082R}
}

@ARTICLE{Kellermann81_compact_radio,
       author = {{Kellermann}, K.~I. and {Pauliny-Toth}, I.~I.~K.},
        title = "{Compact radio sources}",
      journal = {\araa},
         year = 1981,
        month = jan,
       volume = {19},
        pages = {373-410},
          doi = {10.1146/annurev.aa.19.090181.002105},
       adsurl = {https://ui.adsabs.harvard.edu/abs/1981ARA&A..19..373K}
}

@ARTICLE{Chen26_radio_emission,
       author = {{Chen}, Sina and {Behar}, Ehud and {Laor}, Ari and {Arav}, Nahum},
        title = "{Radio Emission from Broad Absorption Line Quasars}",
      journal = {\apj},
         year = 2026,
        month = feb,
       volume = {998},
       number = {2},
          eid = {266},
        pages = {266},
          doi = {10.3847/1538-4357/ae3c7f},
archivePrefix = {arXiv},
       eprint = {2508.21591},
 primaryClass = {astro-ph.GA},
       adsurl = {https://ui.adsabs.harvard.edu/abs/2026ApJ...998..266C}
}

@ARTICLE{Gehrels86_confidence_limits,
       author = {{Gehrels}, N.},
        title = "{Confidence Limits for Small Numbers of Events in Astrophysical Data}",
      journal = {\apj},
         year = 1986,
        month = apr,
       volume = {303},
        pages = {336},
          doi = {10.1086/164079},
       adsurl = {https://ui.adsabs.harvard.edu/abs/1986ApJ...303..336G}
}

@ARTICLE{Temple24_o_iii,
       author = {{Temple}, Matthew J. and {Rankine}, Amy L. and {Banerji}, Manda and {Hennawi}, Joseph F. and {Hewett}, Paul C. and {Matthews}, James H. and {Nanni}, Riccardo and {Ricci}, Claudio and {Richards}, Gordon T.},
        title = "{[O III] emission in z {\ensuremath{\approx}} 2 quasars with and without broad absorption lines}",
      journal = {\mnras},
         year = 2024,
        month = jul,
       volume = {532},
       number = {1},
        pages = {424-437},
          doi = {10.1093/mnras/stae1524},
archivePrefix = {arXiv},
       eprint = {2406.11968},
 primaryClass = {astro-ph.GA},
       adsurl = {https://ui.adsabs.harvard.edu/abs/2024MNRAS.532..424T}
}

@ARTICLE{Xu20_evidence_that,
       author = {{Xu}, Xinfeng and {Zakamska}, Nadia L. and {Arav}, Nahum and {Miller}, Timothy and {Benn}, Chris},
        title = "{Evidence that emission and absorption outflows in quasars are related}",
      journal = {\mnras},
         year = 2020,
        month = jun,
       volume = {495},
       number = {1},
        pages = {305-320},
          doi = {10.1093/mnras/staa1142},
archivePrefix = {arXiv},
       eprint = {1912.09865},
 primaryClass = {astro-ph.GA},
       adsurl = {https://ui.adsabs.harvard.edu/abs/2020MNRAS.495..305X}
}

@ARTICLE{Escott25_unveiling_agn,
       author = {{Escott}, Emmy L. and {Morabito}, Leah K. and {Scholtz}, Jan and {Hickox}, Ryan C. and {Harrison}, Chris M. and {Alexander}, David M. and {Arnaudova}, Marina I. and {Smith}, Daniel J.~B. and {Duncan}, Kenneth J. and {Petley}, James and {Kondapally}, Rohit and {Calistro Rivera}, Gabriela and {Kolwa}, Sthabile},
        title = "{Unveiling AGN outflows: [O III] outflow detection rates and correlation with low-frequency radio emission}",
      journal = {\mnras},
         year = 2025,
        month = jan,
       volume = {536},
       number = {2},
        pages = {1166-1179},
          doi = {10.1093/mnras/stae2645},
archivePrefix = {arXiv},
       eprint = {2411.19326},
 primaryClass = {astro-ph.GA},
       adsurl = {https://ui.adsabs.harvard.edu/abs/2025MNRAS.536.1166E}
}

@ARTICLE{Escott26_sub-arcsecond_ilt,
       author = {{Escott}, Emmy L. and {Morabito}, Leah K. and {Sweijen}, Frits and {Harrison}, Chris M. and {Petley}, James and {de Jong}, Jurjen M.~G.~H.~J. and {van Weeren}, Reinout J. and {Higginson}, Thomas S. and {Prandoni}, Isabella and {Miley}, George and {R{\"o}ttgering}, Huub J.~A.},
        title = "{The sub-arcsecond ILT view of the Bo{\"o}tes Deep Field: a link between low-frequency kiloparsec radio morphology and AGN-driven ionized outflows}",
      journal = {\mnras},
         year = 2026,
        month = apr,
       volume = {547},
       number = {4},
          eid = {stag274},
        pages = {stag274},
          doi = {10.1093/mnras/stag274},
archivePrefix = {arXiv},
       eprint = {2602.05318},
 primaryClass = {astro-ph.GA},
       adsurl = {https://ui.adsabs.harvard.edu/abs/2026MNRAS.547ag274E}
}

%\begin{thebibliography}{99}
%\end{thebibliography}

%%%%%%%%%%%%%%%%%%%%%%%%%%%%%%%%%%%%%%%%%%%%%%%%%%

%%%%%%%%%%%%%%%%% APPENDICES %%%%%%%%%%%%%%%%%%%%%

\appendix

\section[log R distributions]{$\bmath{\log R}$ distributions}
\label{app:logR}
\begin{figure*}
    \centering
    \includegraphics[width=0.8\linewidth]{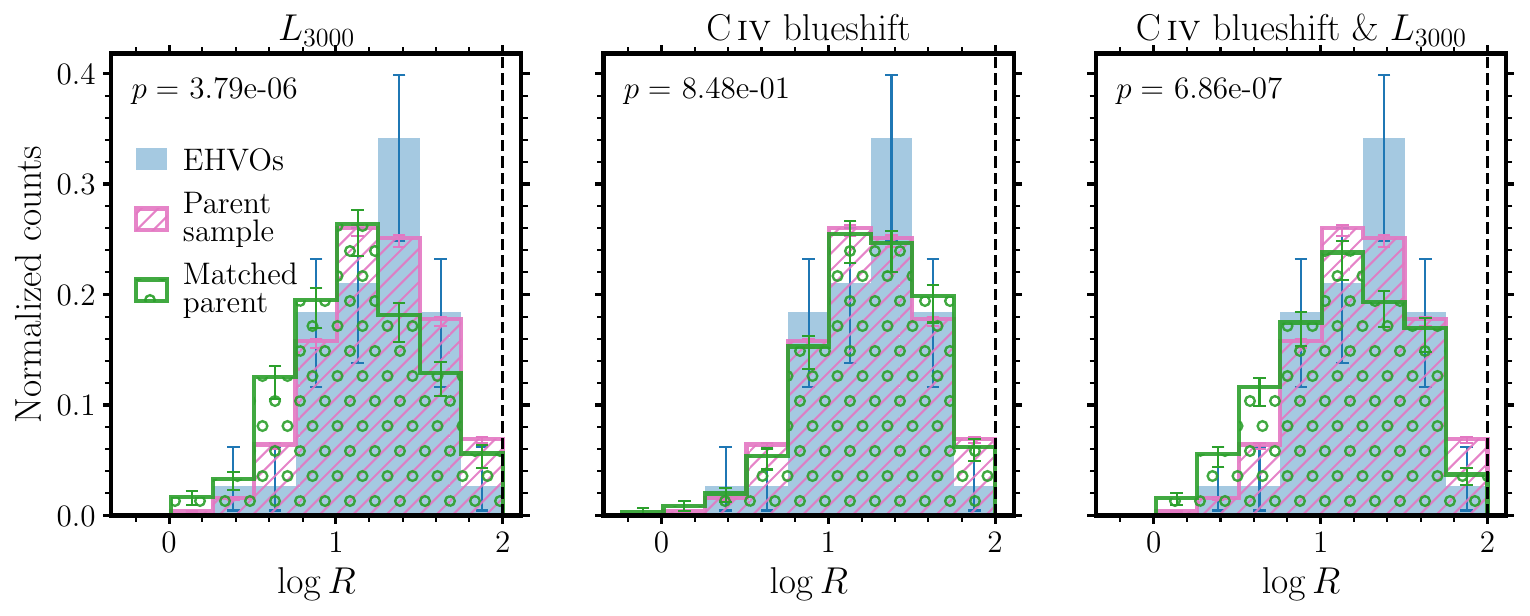}
    \caption{\change{$\log R$ distributions of the EHVO (filled blue histograms),  and parent (pink diagonal hatching) populations of the radio-quiet populations. From left to right, we additionally show the parent sample matched on $L_{3000}$, {\CIV} blueshift, and {\CIV} blueshift and $L_{3000}$ simultaneously (green circles). Errorbars are Poisson uncertainties following Gehrels' method for low number statistics \citep{Gehrels86_confidence_limits}. The $p$-values from KS tests between the matched and parent samples are displayed in the top-left of each panel. Matching does not significantly change the $\log R$ distribution of the parent sample; however, there is a tentative shift towards lower $\log R$ values when the parent sample is matched in $L_{3000}$.}}
    \label{fig:logR_match}
\end{figure*}

\change{We assess if the similarity in the radio-loudness ($\log R$) distributions between the EHVO and parent samples is responsible for the increase in radio-detection fraction when matching on {\CIV} blueshift and/or $L_{3000}$. In Fig.~\ref{fig:logR_match} we show the same radio-quiet parent and EHVO samples' $\log R$ distributions alongside those of the matched samples. The EHVOs are more luminous at 3000\,{\AA} such that matching on $L_{3000}$ produces a matched parent sample that is more optically luminous than the parent sample average. While the $\log R$ distribution is statistically similar to that of the parent sample, there is a tentative shift towards lower radio-loudness values suggesting that similarly optically bright quasars in the parent sample have lower radio luminosities than their EHVO counterparts. A KS test between the parent and matched sample returns a $p\text{-value}\approx3.79\text{e-6}$ suggesting that we can reject the null hypothesis that the matched and parent $\log R$ distributions are the same. Meanwhile, the matching in {\CIV} blueshift produces a $\log R$ distribution that is indistinguishable from the parent sample. Matching on {\CIV} blueshift and $L_{3000}$ simultaneously produces a $\log R$ distribution that is intermediate between those obtained when matching separately.}

%%%%%%%%%%%%%%%%%%%%%%%%%%%%%%%%%%%%%%%%%%%%%%%%%%

% Don't change these lines
\bsp	% typesetting comment
\label{lastpage}
\end{document}